\documentclass[twoside,onecolumn,a4paper,10pt]{revtex4-1}
\def\corref#1{}

\usepackage[T1]{fontenc}
\usepackage{textcomp}
\usepackage{graphicx}
\usepackage{xcolor}
\usepackage{hyperref}
\usepackage{amssymb}
\usepackage{amsmath,mathtools}

\let\oldenumerate\enumerate
\renewcommand{\enumerate}{
  \oldenumerate
  \setlength{\itemsep}{1pt}
  \setlength{\parskip}{0pt}
  \setlength{\parsep}{0pt}
}
\let\olditemize\itemize
\renewcommand{\itemize}{
  \olditemize
  \setlength{\itemsep}{1pt}
  \setlength{\parskip}{0pt}
  \setlength{\parsep}{0pt}
}

\newcommand\colorsout[1]{\bgroup \markoverwith{\textcolor{#1}{\rule[0.5ex]{2pt}{0.4pt}}}\ULon}

\newcommand{\unit}[2]{#1\,#2}
\newcommand{\unitr}[2]{\unit{#1}{\mathrm{#2}}}

\newcommand\sref[1]{Sec.~\ref{#1}}
\newcommand\fref[1]{Fig.~\ref{#1}}
\newcommand\eref[1]{Eq.~\eqref{#1}}

\newcommand\siesta{\textsc{siesta}}
\newcommand\tsiesta{\textsc{transiesta}}
\newcommand\tbtrans{\textsc{tbtrans}}
\newcommand\phtrans{\textsc{phtrans}}
\newcommand\sisl{\textsc{sisl}}
\newcommand\vasp{\textsc{vasp}}
\newcommand\gulp{\textsc{gulp}}

\newcommand{\ie}{\emph{i.\,e.}}

\newcommand\G{\mathbf{G}}

\newcommand\Dyn{\mathbf{D}}
\newcommand\HH{\mathbf{H}}

\newcommand\EDM{\boldsymbol{\mathcal{E}}}
\newcommand\DM{\boldsymbol{\rho}}
\newcommand\DE{\DM_\Eq}
\newcommand\DN{\DM_\Neq}
\newcommand\Spec{\mathcal{A}}

\newcommand\SO{\mathbf{S}}
\newcommand\ncor{\boldsymbol{\Delta}}
\newcommand\SE{\boldsymbol{\Sigma}}
\newcommand\Scat{\boldsymbol{\Gamma}}
\newcommand\rr{\mathbf{r}}
\newcommand\RR{\mathbf{R}}
\newcommand\kk{\mathbf{k}}
\newcommand\qq{\mathbf{q}}
\newcommand\kT{k_BT}
\newcommand\sumE{\sum}%^{\mathfrak{E}}}

\newcommand\idxE{\mathfrak{e}}

\newcommand\cd{\!\dd}
\newcommand\dd{\mathrm{d}}
\newcommand\E{\epsilon}
\newcommand\eig{\varepsilon}
\newcommand\TT{\mathcal{T}}
\newcommand\RE{\mathcal{R}}
\newcommand\wX{\widetilde{\mathbf{X}}}
\newcommand\wY{\widetilde{\mathbf{Y}}}

\newcommand\ID{\mathbf{I}}

\newcommand\BZ{\mathrm{BZ}}
\newcommand\Eq{\mathrm{eq}}
\newcommand\Neq{\mathrm{neq}}

\newcommand\ket[1]{|#1\rangle}
\newcommand\bra[1]{\langle#1|}
\newcommand\bk[2]{\langle#1|#2\rangle}

\newcommand\Csix{\ensuremath{\mathrm{C}_{60}}}

\DeclareMathOperator\Var{Var}
\DeclareMathOperator\Tr{Tr}

\DeclareMathOperator\Res{Res}
\let\Im\relax
\DeclareMathOperator\Im{Im}

\newcommand\dEBZ{\!\!\!\!\!}

\graphicspath{{figs/}}

\newlength\matsize
\def\sq{\rule{\matsize}{\matsize}}

\usepackage{algpseudocode}

\begin{document}

% Re-create affiliations...

\author{Nick Papior}
\email{nickpapior@gmail.com}
\affiliation{Center for Nanostructured Graphene (CNG), Department of Micro- and Nanotechnology (DTU Nanotech), Technical University of Denmark, DK-2800 Kgs. Lyngby, Denmark}

\author{Nicol\'as Lorente}
\email{nicolas\_lorente001@ehu.eus}
\affiliation{Centro de F\'{\i}sica de Materiales CFM/MPC (CSIC-UPV/EHU), Paseo Manuel de Lardizabal 5, E-20018 Donostia - San Sebasti\'an, Spain}

\author{Thomas Frederiksen}%
\email{thomas\_frederiksen@ehu.eus}%
\affiliation{Donostia International Physics Center (DIPC) -- UPV/EHU, E-20018 San Sebasti\'an, Spain}
\affiliation{IKERBASQUE, Basque Foundation for Science, E-48013, Bilbao, Spain}

\author{Alberto Garc\'{\i}a}%
\email{albertog@icmab.es}%
\affiliation{Institut de Ci\`encia de Materials de Barcelona (ICMAB-CSIC), Campus de la UAB, E-08193 Bellaterra, Spain}

\author{Mads Brandbyge}%
\email{mabr@nanotech.dtu.dk}
\affiliation{Center for Nanostructured Graphene (CNG), Department of Micro- and Nanotechnology (DTU Nanotech), Technical University of Denmark, DK-2800 Kgs. Lyngby, Denmark}

\def\ead#1{}
\def\cortext[#1]#2{}
\def\author[#1]#2{}
\def\address[#1]#2{}

\title{Improvements on non-equilibrium and transport Green function techniques:\\
the next-generation \tsiesta}

\author[dtu]{Nick Papior\corref{cor}}%
  \ead{nickpapior@gmail.com}%
  \cortext[cor]{Corresponding author}%
\author[csic,dipc]{Nicol\'as Lorente}%
  \ead{nicolas\_lorente001@ehu.eus}
\author[dipc,ikerbasque]{Thomas Frederiksen}%
  \ead{thomas\_frederiksen@ehu.eus}%
\author[icmab]{Alberto Garc\'{\i}a}%
  \ead{albertog@icmab.es}%
\author[dtu]{Mads Brandbyge}%
  \ead{mabr@nanotech.dtu.dk}

\address[dtu]{Center for Nanostructured Graphene (CNG), Department of Micro- and Nanotechnology (DTU Nanotech), 
Technical University of Denmark, DK-2800 Kgs. Lyngby, Denmark}
\address[csic]{Centro de F\'{\i}sica de Materiales CFM/MPC (CSIC-UPV/EHU),
Paseo Manuel de Lardizabal 5, E-20018 Donostia - San Sebasti\'an, Spain}
\address[dipc]{Donostia International Physics Center (DIPC) -- UPV/EHU,
	E-20018 San Sebasti\'an, Spain}
\address[ikerbasque]{IKERBASQUE, Basque Foundation for Science, E-48013, Bilbao, Spain}
\address[icmab]{Institut de Ci\`encia de Materials de Barcelona (ICMAB-CSIC), Campus de la UAB, E-08193 Bellaterra, Spain}

% maketitle creates the abstract box
\begin{abstract}
  We present novel methods implemented within the non-equilibrium Green function
  code (NEGF) {\tsiesta} based on density functional theory (DFT). 
  Our flexible, next-generation DFT-NEGF code handles devices with one or multiple
  electrodes ($N_\idxE\ge1$) with individual chemical potentials and electronic
  temperatures.  We describe its novel methods for electrostatic gating, contour
  optimizations, and assertion of charge conservation, as well as the newly implemented
  algorithms for optimized and scalable matrix inversion, performance-critical pivoting,
  and hybrid parallellization.
  Additionally, a generic NEGF ``post-processing'' code (\tbtrans/\phtrans) for electron
  and phonon transport is presented with several novelties such as Hamiltonian
  interpolations, $N_\idxE\ge1$ electrode capability, bond-currents, generalized interface
  for user-defined tight-binding transport, transmission projection using eigenstates
  of a projected Hamiltonian, and fast inversion algorithms for large-scale simulations
  easily exceeding $10^6$ atoms on workstation computers.
  The new features of both codes are demonstrated and bench-marked for relevant test systems.
\end{abstract}

\maketitle

\section{Introduction}
\label{sec:intro}
The transport of charge, magnetic moments and, in general, any sort of excitation is a
fascinating fundamental physical problem that has demanded attention for a long time~\cite{Feynman}. Today, the interest is enhanced by the technological
needs of an industry increasingly based on devices whose detailed atomistic structure
matters~\cite{SRM}, but the treatment of transport is still a formidable open task.  Spurred
by the fast developments of the microelectronic industry, the first attempts to understand
electronic transport at the atomic scale where based on scattering
theory~\cite{Landauer}. The electron transmission between two semi-infinite reservoirs was
treated in a time-independent fashion solving the scattering matrix connecting the
reservoirs.  At this stage, transport was described as one-electron scattering by a
static contact region and this granted access to many concepts and to devising new
experiments~\cite{christian0,Datta,sanvito}. However, the problem is fundamentally a non-equilibrium one
that requires evolving many-body states
\cite{Delaney,pierre,Mirjani11prb,herve}.

%A quantum leap came with the advent of density functional theory (DFT). 
Density functional theory (DFT) has been one method to address some aspects of this problem.
Conceptually, DFT is a mean-field many-body theory of the ground state. As such, it can in principle
give exact results for the linear conductance because the linear response is a property of
the ground state~\cite{Martin}. Beyond linear conductance, not even \emph{ideal} DFT
works because of the need to describe excited states and dynamics of the system. Such limitations may be mitigated by using time-dependent
DFT \cite{Stefanucci2004,Ventra2004},
%In order to render DFT a practical theory, approximations must be undertaken.
%
%In its present use, DFT is a mean-field theory that permits us to continue with a
%one-electron description but that is of increasing quantitative accuracy as sophisticated
%functionals are implemented.
%
but going beyond the linear regime is highly nontrivial. A main
issue of a DFT description stems from the approximations made
% have a working method 
to compute the ground state. Indeed, it has been recently shown that cases where strong
correlations rein, such as the Coulomb blockade regime, the commonly used exchange-and-correlation functionals fail and new ones have to be
used~\cite{stefanucci}.

Probably the most significant conceptual and practical problem comes from the use of the Kohn-Sham
electronic structure as the working basis for transport calculations~\cite{kurth}. While this
has many limitations and restrictions~\cite{stefanucci,kurth,Jos,Ferdinand,Mera} it currently seems to be the most practical way of obtaining insight based on 
atomistic modeling \cite{Ferdinand,Mera}. For the range of systems where DFT is thought to be of quantitative value, efficient and accurate codes based on a combination of DFT and non-equilibrium Green function
(NEGF) theory have been implemented. The collection of such DFT-NEGF codes is an ever growing
list~\cite{Taylor2001,Brandbyge2002,juanjo,stefan,jaime,aran,fabian,Pecchia2008,Saha2009,Ozaki2010,Chen2012,Bagrets}, but few are multi-functional in the sense of having predictive power for a
variety of physical properties (electrical and heat conductivity, influence of heat
dissipation, etc.) and flexible enough to describe realistic experimental situations (e.g.,
complicated chemical compounds, multi-terminal setups, and devices involving thousands of atoms). 
% Let us take the case of multi-terminal junction transport. Initial
% implementations built on the scattering approach permitted to rationalize and identify
% many physical situations \comment{(can we be more specific?)}~\cite{christian1}.  Yet, it
% is fairly rare to find multi-terminal capable numerical
% codes~\cite{Saha2009,christian2,jaime1,Cook2011}. \comment{(Shouldn't we limit the
%     discussion to DFT-NEGF?)}

% Several codes implements equilibrium Green function (EGF) and non-equilibrium (NEGF)
% techniques \cite{Brandbyge2002,Saha2009,Ozaki2010,Chen2012}. While a few codes allow the
% calculation of multi-terminal junction transport \cite{Saha2009,Cook2011}.

In the present work, we report on a complete rewrite of the \tsiesta\ DFT-NEGF code. An emphasis has been put in
increasing both the efficiency and the accuracy of the calculations. The new \tsiesta\
presents: 1) a huge performance increase using advanced inversion algorithms on top of
efficient threading, 2) an efficient generalization of equations for multi-terminal
systems, 3) a new treatment of thermoelectric effects by allowing temperature gradients, 4)
new gate methods in conjunction with improved electrostatic effects, 5) new contour
integration optimizations for improved convergence, and 6) a fully flexible tight-binding
functionality using Python as back-end.

% While the equations for multi-terminal NEGF calculations already exists,
% \cite{Saha2009,Chen2012}, we note an important thing for reducing the computational
% complexity if 2 or more terminals have the same electronic distribution.
% 
% < section about outlying for the paper, sec X tells this....>

%This paper consists of 3 different sections, all readable in their own entity but thought as parts of a full ensemble. 
The paper is organized as follows:
Section \ref{sec:negf} is devoted to the general
framework of a multi-terminal formulation within DFT-NEGF while section
\ref{sec:implementation} deals with the implementations specific to \tsiesta.
% In \sref{sec:siesta} we describe a few \siesta\
% implementations that deal with generic optimizations of the \siesta\ code. \sref{sec:negf}
% discuss the new NEGF code \tsiesta\ which retains the original implementation's name but
% with added above functionalities, namely multiterminal capabilities, inclusion of gates
% and enhanced electrostatic functionalities, and realistic thermoelectric effects. 
In \sref{sec:tbtrans} we finally cover a generic ``post-processing'' NEGF code
(\tbtrans/\phtrans) to compute, among other features, electron and phonon transmissions
with inputs from a DFT-NEGF description (i.e., \tsiesta\ or similar software) or simply
from some user-supplied tight-binding parameters.
%
%The efficiency of this hybrid parallelized code enables calculations of ``tight-binding''
%Hamiltonians with excess of $10^6$ orbitals on regular workstation machines. This
%flexibility enables reproducing experimental situations or uncovering new
%physics without jeopardizing the numerical accuracy.

\section{Green function theory}
\label{sec:negf}

%\comment{AG: Green function or Green's function? See
%http://www.nature.com/nphys/journal/v2/n10/full/nphys411.html}
% Indeed that read is interesting (I also passed it to Thomas ;)
% I much prefer Green, Green's is tedious to write... :)
The central aim of DFT-NEGF is to obtain a self-consistent description of the electron
density $\DM$ and the effective Kohn-Sham Hamiltonian $\HH$ for an open quantum system
coupled to one or more electrodes. These electrodes are thought to be large enough to be
unperturbed by the presence of electronic currents passing through the scattering region,
i.e. that they can be considered in local equilibrium.  However, if the electrodes are not
in equilibrium with each other, the central part (system) will acquire a non-equilibrium
electron density. In contrast to ground-state DFT, where the electron density is simply
obtained by filling the Kohn-Sham states up to the Fermi level, such simple relation
between occupations and states is not available in the non-equilibrium situation. Instead one
can resort to Green function techniques as outlined below for the steady-state solution.

\begin{figure}
  \centering
  \includegraphics{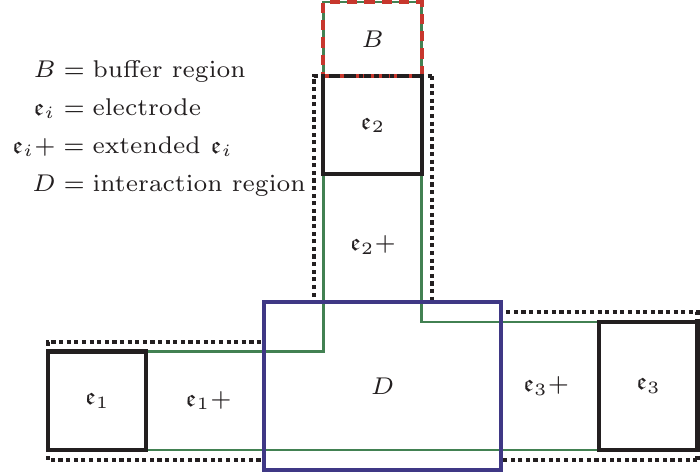}
  \caption{Conceptual system setup for a 3-electrode ($N_\idxE=3$) example. The electrode regions are denoted by $\idxE_i$ (black blocks) and the associated
electrode screening regions by $\idxE_i+$. The scattering/device region is indicated by
      $D$ (blue block). An additional buffer region $B$ (red block) denotes a region removed from the NEGF algorithm. All blocks as a whole represent the supercell used for a \tsiesta\ calculation.
      \label{fig:system-setup}
  }
\end{figure}

% Before we discuss the implementation details of the new \tsiesta, we outline the basic
% equations governing a NEGF treatment of open quantum systems. 
The specifics governing the underlying methodology (\siesta) for atomic-like basis-sets
can be found elsewhere \cite{Soler2002,Brandbyge2002}. Figure \ref{fig:system-setup}
illustrates the kind of generic multi-electrode NEGF setup we have in mind in the
remainder of the paper.
For setups where periodic boundary conditions with given lattice vectors
$\mathbf R$, we apply Bloch $\kk$-point sampling possible for both the DFT-NEGF
self-consistent calculation and the subsequent transport calculation. To keep clarity, we
explicitly add the $\kk$ dependence to the equations while $\idxE$ refers to an electrode
index. Furthermore we write all equations generically with $N_\idxE$ electrodes to clarify
specifics related to any number of electrodes; $N_\idxE\ge 1$. The following expressions
are used throughout the paper
\begin{align}
  \label{eq:Gf}
  \G_\kk(z)&=\Big[z\SO_\kk - \HH_\kk - \sumE_\idxE\SE_{\idxE,\kk}(z)\Big]^{-1},
  \quad \text{ with $z\equiv\E+i\eta$,}
%  \\
%  \label{eq:SEtot}
%  \SE_{\kk}(z) &= \sumE_\idxE\SE_{\idxE,\kk}(z),
  \\
  \Scat_{\idxE,\kk}(z) &=i\big[\SE_{\idxE,\kk}(z)-\SE^\dagger_{\idxE,\kk}(z)\big],
  \\
  \Spec_{\idxE,\kk}(z) &=\G_\kk(z)\Scat_{\idxE,\kk}(z)\G_\kk^\dagger(z),
  \\
  % \label{eq:Spec}
  % \Spec_{\kk}(z) &=\sumE_\idxE\Spec_{\idxE,\kk}(z) = i[\G_\kk(z)-\G_\kk^\dagger(z)]
  % \\
  \label{eq:DM}
  \DM % &=  \iint_\BZ\dEBZ\cd \kk\dd \E\,\G^<_\kk(\E)e^{i\kk\cdot\RR} % TF: G< not used elsewhere
  %&\equiv  \iint_\BZ\dEBZ\cd \kk \dd\E\, \G_\kk (\E)\Scat_{\idxE,\kk}(\E)n_{F,\idxE}(\E)   \G^\dagger_\kk(\E) e^{i\kk\cdot\RR},
  &= \frac1{2\pi}\iint_\BZ\dEBZ\cd \kk \dd\E\,\sumE_\idxE \Spec_{\idxE,\kk} (z) n_{F,\idxE}(\E)
  e^{-i\kk\cdot\RR},%\equiv\sum_\idxE\delta\DM^\idxE,
\end{align}
where $\G_\kk$/$\G_\kk^\dagger$ is the retarded/advanced Green function at energy $\E$
(with a small positive constant $\eta=0^+$), and $\HH_\kk=\HH e^{i\kk\cdot\RR}$,
$\SO_\kk=\SO e^{i\kk\cdot\RR}$, the Hamiltonian and overlap matrix at $\kk$ in the
scattering region with $\RR$ being a lattice vector in the periodic directions. The
self-energy and spectral function of electrode $\idxE$ are $\SE_{\idxE}$ and
$\Spec_\idxE$, respectively, with associated broadening matrix $\Scat_\idxE$. Lastly,
$\DM$ is the non-equilibrium density matrix.
BZ denotes here, and in the following, the Brillouin zone \emph{average} (i.e., it
includes a normalization corresponding to the appropriate Brillouin zone
volume). Equation~\eqref{eq:DM} is the density matrix for equilibrium \emph{and}
non-equilibrium (disregarding bound states).
We require a Hermitian Hamiltonian and express the chemical potential as $\mu$, and the
temperature as $\kT$. A combined quantity is defined $\varsigma\equiv\{\mu,\kT\}$. We will
freely denote a Fermi distribution by $n_{F,\varsigma}$ as well as $n_{F,\idxE}$ where the
latter implicitly refers to $\varsigma$ belonging to the electrode $\idxE$.
\tsiesta\ is also implemented with spin-polarization and we will omit the
factor of $2$ for non-polarized calculations, thus equations are for one
spin-channel, unless otherwise stated.
Lastly, we omit using the so-called ``transport direction'' which is ill-defined for
nonparallel electrodes. As such our implementation of \tsiesta\ only deals with the
semi-infinite directions of each electrode. This is apparent in $N_\idxE>2$ calculations
as performed in Refs.~\cite{Palsgaard2015,Jacobsen2016}.

Finally, we define another central quantity for the Green function technique, namely the energy density matrix
$\EDM$, as
\begin{equation}
  \label{eq:EDM}
  \EDM
  = \frac1{2\pi}\iint_\BZ\dEBZ\cd \kk \dd\E\,\E \sumE_\idxE \Spec_{\idxE,\kk} (z) n_{F,\idxE}(\E)
  e^{-i\kk\cdot\RR},
\end{equation}
which enables force calculations under non-equilibrium situations \cite{DiVentra2000,Brandbyge2003a}.

\subsection{Equilibrium (EGF)}
\label{ssec:egf}
In (global) equilibrium all Fermi distribution functions are equal, i.e.,
$n_{F,\idxE}(\E)=n_F(\E)$, and \eref{eq:DM} can be reduced to
% \begin{align}
%   \label{eq:equi}
%   \DE &= \iint_\BZ\dEBZ\cd \kk \dd\E\, %
%   %\G_\kk \sumE_\idxE\Scat_{\idxE,\kk} \G^\dagger_\kk n_F(\E) %
%   \sumE_\idxE\Spec_{\idxE,\kk} (z) n_F(\E)e^{-i\kk\cdot\RR}
%   \\
% %   &= i\iint_\BZ\dEBZ\cd \kk\dd\E\, %
% %   \G_\kk (z)\big[\SE_{\kk}(z)-\SE^\dagger_{\kk}(z) \big]
% %   \G^\dagger_\kk (z) n_F(\E)e^{-i\kk\cdot\RR} %
% %   \\
% % %  \shortintertext{by introducing \eref{eq:Gf}}
% %  & = i\iint_\BZ\dEBZ\cd \kk\dd\E\, %
% %   \G_\kk(z)  \left[%
% %     \G^{\dagger,-1}_\kk(z)  - \G^{-1}_\kk(z)  + 2i\eta\SO_\kk %
% %   \right]\G^\dagger_\kk(z)  n_F(\E)e^{-i\kk\cdot\RR}
% %   \\
%   & = i\iint_\BZ\dEBZ\cd \kk\dd\E%
%     \left[\G_\kk(z)  - \G^\dagger_\kk (z)+ 2i\eta\G_\kk(z)\SO_\kk\G^\dagger_\kk(z) %
%     \right]n_F(\E)e^{-i\kk\cdot\RR}
%   \\
%   \label{eq:egf:final}
%   &
%   \approxeq i\iint_\BZ\dEBZ\cd \kk\dd\E%
%     \left[\G_\kk(z)   - \G^\dagger_\kk(z)  \right]n_F(\E)e^{-i\kk\cdot\RR}.
% \end{align} %
% %We define \eref{eq:egf:final} as the equilibrium density $\DE$ belonging to the chemical
% %potential $n_{F}$. As $\eta\to0^+$ the last term vanishes.
% In the last step the term proportional to $\eta$ vanishes as $\eta\to0^+$.
% \comment{Is this derivation really necessary? Can't we just use the standard expression $\Spec_\kk(z)=i\left[\G_\kk(z) - \G^\dagger_\kk(z)  \right]=\sumE_\idxE\Spec_{\idxE,\kk} (z)$, now inserted as \eref{eq:Spec} and simply write}
\begin{equation}
  \label{eq:equi}
  \DE = \frac i{2\pi}\iint_\BZ\dEBZ\cd \kk\dd\E%
    \left[\G_\kk(z)   - \G^\dagger_\kk(z)  \right]n_F(\E)e^{-i\kk\cdot\RR}.
\end{equation} %

To circumvent the meticulous and tedious integration in \eref{eq:equi} along the energy
axis, it is advantageous to use the residue theorem \cite{Brandbyge2002}. The advantage
is that the Green function, which varies quickly near the poles on the real axis, is
analytic and much smoother in the complex plane, which in turn allows for numerically
accurate quadrature methods. An example of the smoothing in the complex plane can be found
in the supplementary material (SM). As the Green function has poles on the real axis (the
eigenvalues of its inverse) and the Fermi function $n_F$ has poles at
$z_\nu=i\kT\pi(2\nu+1)$ with $\Res n_F(z_\nu)=\kT$ for $\nu\in\mathbb{N}$, we have
according to the residue theorem that
\begin{equation}
  \label{eq:egf:res}
  \oint\dd z \big[ \G_\kk(z)-\G_\kk^\dagger(z) \big] n_F(z) = -2\pi i \kT\sum_{z_\nu}
  \big[ \G_\kk(z_\nu)-\G_\kk^\dagger(z_\nu) \big],
\end{equation}
which follows if $z\in\mathbb{R} +i\eta$, $\eta\to0^+$.

\begin{figure}
  \centering%
  \includegraphics[clip,trim=.2cm 0 1cm 1.6cm]{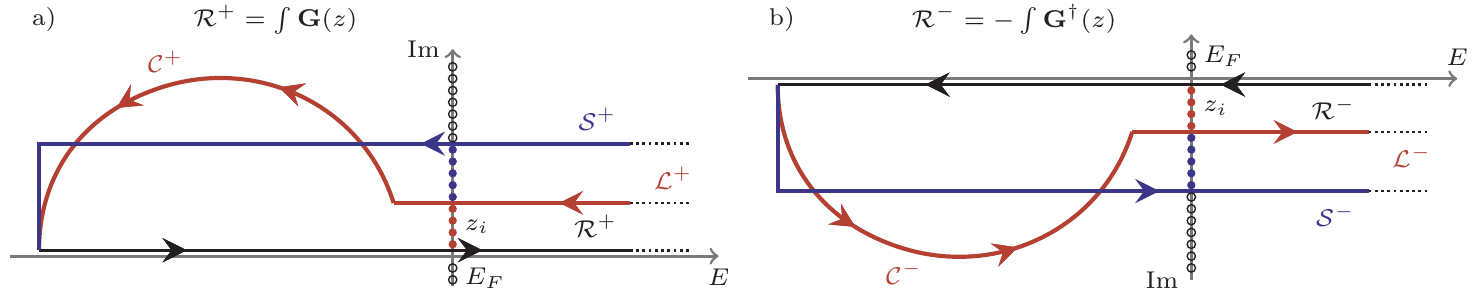} %
  \caption{Two enclosing contours in the complex plane. The red contour is a \emph{circle}
      contour while a square contour is shown by the blue line. Both are equivalent
      mathematically. %
      The arrows indicate the direction of the contour integration. a) is the integration
      of the retarded Green function while b) is the advanced Green function. Note the
      sign change of the advanced Green function which results in an opposite direction integration.
      \label{fig:res}
  }
\end{figure}

An example of two different, but mathematically equivalent, contours are shown in
\fref{fig:res}a. To calculate the real axis integral one divides the enclosed contour into
the integral along the real axis and the remaining of the contour. We note that
$n_F(z)\to0$ for $\Re z\gg E_F$ which avoids the need for a fully enclosed contour as
$\lim_{z\to\infty+i\eta}\int_{\mathcal R_{\mathrm{up}}}^{(\mathcal{S/L})_\mathrm{up}}\dd z
f(z)n_F(z)\to0$.
We stress that \emph{all} enclosed contours in the lower/upper complex plane are
mathematically equivalent, as long as the lower bound is below the lowest eigenvalue in
the Brillouin zone.
%
%The careful reader will notice that 
The residue theorem can be applied two times for
$\G_\kk(z)-\G_\kk^\dagger(z)$: We use the positive
part of the imaginary coordinate system with $\Im z>0$ for integrating
$\G_\kk(z)$, and the negative part of the imaginary coordinate system with
$\Im z<0$ for integrating $\G_\kk^\dagger(z)$. This is indicated in
\fref{fig:res}a/b with $\mathcal R^{+/-}$, respectively.
Importantly, the imaginary part of the line contour $\mathcal L/\mathcal S$ should be
chosen large enough so that the Green function indeed is smooth. A higher number of poles
increases the distance to the real axis. Thus one should take care of the number of poles
used in the calculation as the imaginary part is solely determined by the temperature. For
10 poles and a temperature of $\unitr{25}{meV}$ the imaginary part becomes
$\sim\unitr{1.57}{eV}$. We emphasize that to ensure a consistent interpretation,
irrespective of the electronic temperature, it is better to derive the number of poles from a fixed
energy scale on the imaginary axis rather than choosing it directly.
Furthermore, it is required that the lower bound of the contour integration is well below
the lowest eigenvalue of the system as the Green function fans out when increasing the
complex energy. See the SM for an interactive illustration of these points.

\subsection{Non-equilibrium (NEGF)}
\label{ssec:negf}

Non-equilibrium arises due to differences between the electrode electronic distributions
via $\varsigma_\idxE\neq\varsigma_{\idxE'}$. That is, either a chemical potential
difference, an electronic temperature difference, or a combination of these. We define the
bias window with a lower/upper bound as
$\operatorname{min}(\mu_\idxE)$/$\operatorname{max}(\mu_\idxE)$ with appropriate tails of
the Fermi functions.
%The NEGF part is the most time consuming part as it requires the Green function \emph{and} a triple matrix product.
%
Starting from \eref{eq:DM} and adding
\begin{gather} 
%  \label{eq:neq}
%  \DM = \frac1{2\pi}\iint_\BZ\dEBZ\cd \kk\dd\E\, %
%  \G_\kk \sumE_\idxE\Scat_{\idxE,\kk}n_{F,\idxE}(\E) \G^\dagger_\kk e^{-i\kk\cdot\RR}%
%  \\
  \label{eq:neq:add}
  0=(1-1)\sumE_{\idxE'\neq\idxE}\Spec_{\idxE',\kk}(z)n_{F,\idxE}(\E) e^{-i\kk\cdot\RR},
\end{gather}
%Adding \eref{eq:neq} and \eref{eq:neq:add} and reducing
(note that the products $\Spec_{\idxE',\kk}(z)n_{F,\idxE}(\E)$ refer to different electrodes)
we can write the density matrix as
\begin{align}
%  \DM &= \DM^\idxE+\sumE_{\idxE'\neq\idxE}\iint_\BZ\dEBZ\cd \kk\dd\E\, %
%  \G_\kk \Scat_{\idxE',\kk}
%  \G^\dagger_\kk e^{-i\kk\cdot\RR}\big[n_{F,\idxE'}(\E)-n_{F,\idxE}(\E)\big], %
%  \\
  \label{eq:neq:elec}
  \DM &= \DE^\idxE+\sumE_{\idxE'\neq\idxE}\ncor_{\idxE'}^\idxE\equiv \DN^\idxE ,\\
  \label{eq:neq:elec1}
  \DE^\idxE &\equiv  \frac i{2\pi}\iint_\BZ\dEBZ\cd \kk\dd\E%
  \left[\G_\kk(z)   - \G^\dagger_\kk(z)  \right]n_{F,\idxE}(\E)e^{-i\kk\cdot\RR}, \\
  \label{eq:neq:elec2}
  \ncor_{\idxE'}^\idxE &\equiv
  \frac1{2\pi}\iint_\BZ\dEBZ\cd \kk\dd\E\, 
  \Spec_{\idxE',\kk} (z) e^{-i\kk\cdot\RR}\big[n_{F,\idxE'}(\E)-n_{F,\idxE}(\E)\big],
\end{align}
where we call $\ncor_{\idxE'}^\idxE$ a non-equilibrium correction term for the equilibrium
density of electrode $\idxE$ due to electrode $\idxE'$. This reduces the real axis
integral to be confined in the bias window with respect to the different Fermi
distributions. Equation~\eqref{eq:neq:elec} deserves a few comments. It can be expressed
equivalently for all electrodes $\idxE$, and thus one finds $N_\idxE$ different
expressions for the same density $\DM=\DN^\idxE=\DN^{\idxE'}=\cdots$. If two or more
electrodes have the same Fermi distribution, $\varsigma_{\idxE}=\varsigma_{\idxE'}$, we
find $\DE^{\idxE}=\DE^{\idxE'}$ and $\ncor_{\idxE'}^{\idxE}=0$, thus we can reduce
$N_\idxE$ to $N_\varsigma$ different expressions. So for any $N_\idxE>2$ electrodes with
$2$ different Fermi distributions we only have $2$ equations with different terms
(although the two equations are mathematically equivalent). We stress that the number of
correction terms $\ncor_{\idxE'}^\idxE$ for each electrode depends on the number of
equivalent Fermi distributions.
Equivalently, \eref{eq:neq:elec} can be written more compactly as
\begin{equation}
  \label{eq:neq:mu}
  \DM
  %=\DE^\varsigma+\frac1{2\pi}\sumE_{\idxE|\varsigma_\idxE\neq\varsigma}\iint_\BZ\dEBZ\cd \kk\dd\E\, %
  %\G_\kk \Scat_{\idxE,\kk}
  %\G^\dagger_\kk e^{-i\kk\cdot\RR}\big[n_{F,\varsigma_\idxE}(\E)-n_{F,\varsigma}(\E)\big]
  =
  \DE^\varsigma+\sumE_{\mathclap{\idxE|\varsigma_\idxE\neq\varsigma}}\ncor_{\idxE}^\varsigma\equiv \DN^\varsigma,
\end{equation}
where $\idxE|\varsigma_\idxE\neq\varsigma$ are electrodes with Fermi distributions
different from $\varsigma$. Equation~\eqref{eq:neq:elec} is equivalent to \eref{eq:neq:mu} where the
former have possible duplicates and the latter does not.
These considerations also apply to the (non-equilibrium) energy density matrix, \eref{eq:EDM},
in a similar manner.

Compared to the equilibrium case (\sref{ssec:egf}) we note that the non-equilibrium case
is numerically more demanding in terms of matrix operations as the calculation of $\DM$,
in addition to the inversion for $\G_\kk(z)$ needed in \eref{eq:equi} and
\eref{eq:neq:elec1}, also requires the evaluation of triple matrix products for
$\Spec_{\idxE,\kk}$, as seen in \eref{eq:neq:elec2}.

\section{Implementation details in \texorpdfstring{\tsiesta}{TranSIESTA}}
\label{sec:implementation}

\subsection{Complex contour optimization}
\label{sec:contour}

The equilibrium contour in NEGF calculations can be chosen from a range of different
shapes and methods to integrate. 
Here we describe the nontrivial task of selecting an optimum equilibrium contour. We have 
implemented several different methods, from Newton-Cotes to advanced quadrature methods,
using Legendre polynomials or Tanh-Sinh quadrature \cite{Takahasi1974}. Both
circle and square contours are possible. Furthermore the continued fraction method
suggested by Ozaki \cite{Ozaki2010} is also implemented. Its strength is that it has only
\emph{one} convergence parameter, which is the number of poles, whereas the quadrature methods have (at
least) three convergence parameters(number of poles ($z_i$), points on the line $\mathcal L^+$, and points on
circle/square $\mathcal C^+/\mathcal S^+$).

% Even though the calculation of the non-equilibrium contour is the most time consuming,
% optimizing the equilibrium contour is also of importance. In \tsiesta\ we deal with
% $N_\varsigma$ different equilibrium contours to approximate the density with the highest
% accuracy. 

% In the following we will concentrate on the lower part of the equilibrium integration
% contour. First we note that the Green function becomes zero far from the band-bottom of
% the electronic structure. Thus for low $\E$ the integration is $0$. We utilize a circle
% contour in the complex plane as shown in \fref{fig:res}. In general a Gaussian quadrature
% method using the Legendre polynomials are used. We also implement the Tanh-Sinh quadrature
% method \cite{Takahasi1974} (and regular Newton-Cotes integration for full range
% consistency). Yet the Legendre quadrature outperforms all tried methods. 
%
A novel selection of quadrature points in $\mathcal C^+/\mathcal S^+$ can be realized by
examining \fref{fig:res}. It is evident that the two contours $\mathcal C^{+/-}$ for the
retarded and advanced Green function add up to a connected circle. Hence we can
consider them as \emph{one} integration path and choose the quadrature to span the entire
$\mathcal C^++\mathcal C^-$ contour. In practice one only chooses the \emph{right} half of
the abscissa on the $\mathcal C^++\mathcal C^-$ contour and effectively one uses a half
quadrature on the $\mathcal C^+$ contour. We will denote this as the \emph{right-side}
scheme. 
% will chunk points close to the $\mathcal L^{+/-}$ contour which should increase precision
% where the contour is closer to the real axis (less smooth). 
This trick allows one to slightly reduce the number of equilibrium contour
points without loss of accuracy\footnote{This is due to the constant DOS for the lower energy part
    of the contour where the circle is far below the lowest lying eigenvalue and far in
    the complex plane.}.

\begin{figure}
  \centering
  \includegraphics{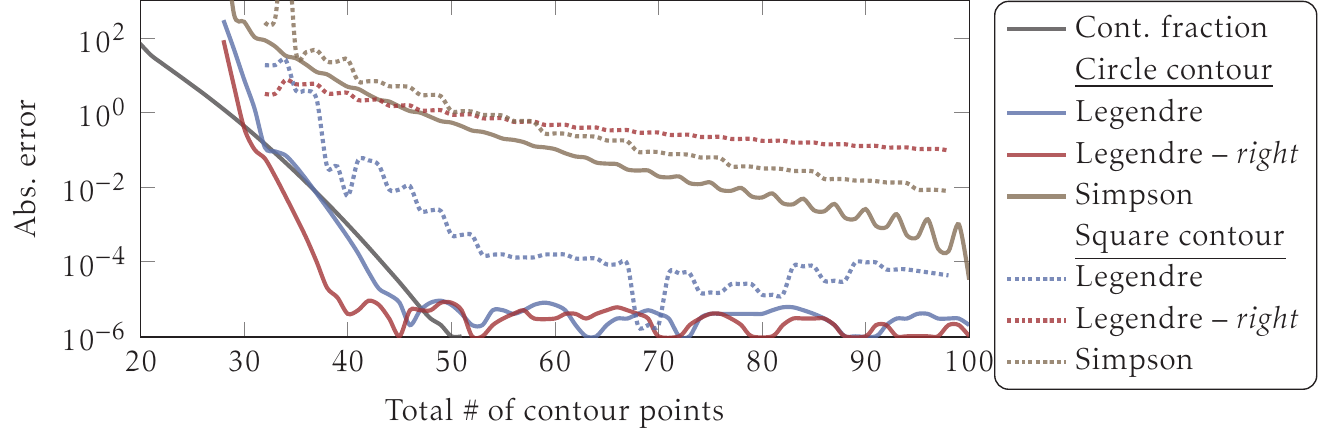}
  \caption{Test calculation on a metallic one-dimensional gold chain using Gaussian quadrature methods. Comparison of the square contour vs.~the circle
      contour using the variants of integration methods: Regular Legendre, 
 Legendre-\emph{right-side} and Simpson quadrature. The Legendre-\emph{right-side} seems better or at
      least on-par with the regular Legendre quadrature on the circle contour.
      % \revision{TF: Add \emph{side} in legend?}
      % Sadly, I forgot the data files for this plot at my workstation at DTU.
      % I should be able to retrieve it at some point :)
      % TODO
      \label{fig:gauss}
  }
\end{figure}

In order to illustrate the convergence properties of the equilibrium contour we have
investigated a two-electrode gold (slab) system which is connected via a
one-dimensional (1D) chain and calculated the free energy as a function of contour
points. As the convergence path is non-deterministic and the ``correct'' value cannot be
found we define the error against the free energy calculated with 300 energy points on the
respective contour. As such the reference is itself. We stress that this study is
difficult to extrapolate to arbitrary systems, yet it can be indicative of the convergence
properties for the different quadrature methods and illustrates how critical the choice of
method can be. The results are seen in \fref{fig:gauss}. The circle and square quadratures
have $16$ poles and the circle uses $10$ $\mathcal L^+$ points. Both the circle and square
contours are presented using both the standard and the \emph{right-side} scheme as well as the
continued fraction method. Note that the numerical accuracy limits the error to $10^{-6}$.

We see that the circle contour benefits from the \emph{right-side} scheme which  performs the best in this setup. On the contrary, the square contour does not benefit from the \emph{right-side} scheme. Furthermore, we see a slow convergence of the Simpson method (order 3 of Newton-Cotes method).
Lastly, the continued fraction scheme \cite{Ozaki2010}  converges fast and indeed is a powerful method due to its simplicity.
We stress that changing the initial $\mathcal L^+$ and/or number of poles will change the
convergence properties of the shown methods.

%  We do this for both the circle and the square
% contour from a clean calculation with an equal amount of poles. As the comparison is with
% itself it is difficult to compare across the schemes. Nevertheless we can see that the
% convergence for the Simpson method is poorer than the quadrature methods. The circle
% contours outperforms the square contour for few points and it may seem that for high
% number of points the square seems best. Nevertheless are the \emph{right-side} methods just as
% good as their regular alternatives.

\subsection{Weighing \texorpdfstring{$\DM$}{rho} and bound states}
\label{ssec:weight}

As shown in \sref{ssec:negf} several different expressions exist for the non-equilibrium
% * <nicolas_lorente001@ehu.eus> 2016-07-13T16:15:47.536Z:
%
% just a silly remark: "to weigh" does not take a "t"
%
% ^.
density matrix $\DN$, depending on the choice of electrode for the equilibrium part in
\eref{eq:neq:elec} and \eref{eq:neq:mu}.  Under non-equilibrium conditions these
expressions numerically yield different densities, in particular if bound states are
present.
Per definition bound states do not couple to
any of the electrodes via the spectral function, i.e.,
$\langle\psi_{\mathrm{bound}}|\Spec_\idxE|\psi_{\mathrm{bound}}\rangle=0$. 
Their contributions to the non-equilibrium density $\DN$ thus only derive from \eref{eq:neq:elec1} -- but never from \eref{eq:neq:elec2} -- as bound states are included in the electronic spectrum of $\G_\kk(z)$.
The filling level of bound states thus depends on the choice of equilibrium electrode in \eref{eq:neq:elec1}.

To avoid the arbitrariness of selecting one equilibrium electrode, and to reduce numerical
errors, the physical quantity $\DN$ is expressed as an appropriate average over each of
the numerically unequal expressions
\begin{equation}
  \label{eq:weighingDM}
  \DN = \sumE_\idxE w_\idxE \DN^\idxE,
\end{equation}
where $w_\idxE$ is an appropriately chosen weight function satisfying
$\sum_\idxE w_\idxE =1$. Several DFT-NEGF implementations apply such a weighing scheme,
but with differences in the particular choice of weights $w_\idxE$
\cite{Brandbyge2002,Li2007,Saha2009}.
%
% As an example we imagine a system with a bound state at the Fermi level. Applying a bias
% between two electrodes ($L/R$) will create two seemingly equivalent densities. However,
% $\Spec_L=\Spec_R=0$ and hence only the equilibrium part of the density contributes to the
% total density (the Green function captures all states). In the low chemical potential
% equilibrium density the bound state is not captured, whereas for the higher chemical
% potential, the equilibrium density contains the bound state. As the bound state lies in
% the bias window it becomes unclear how one should populate it. Should it 1) retain its
% equilibrium population, 2) be empty, thus populate it according to the low chemical
% potential or 3) occupy the state according to the highest chemical potential or 4) some
% fraction of each. In \tsiesta\ we choose an equal amount of each part.
%
% \cite{Li2007} also deals with this problem and populate it according to the geometric mean
% of the trace of the atomic weights.
%
% Locating bound states can be performed using \tbtrans\ by calculating the DOS from the
% spectral functions as well as from the Green function.
%
%An example of a bound state is given in the supplementary material.
%
% To \tsiesta\
% calculates all \eref{eq:neq:mu} and takes an average to approximate the true
% quantity.
%
%
We extend the argumentation of Ref.~\cite{Brandbyge2002} for the weighing to
a multi-terminal expression, and find the weights that minimize the variance of the final
density to be
\begin{gather}
  \boldsymbol\theta_\idxE = \sum_{\idxE'\neq\idxE}\Var[\ncor^{\idxE}_{\idxE'}],
  \\
  \label{eq:weight:non-eq:solution}
  % Full complex equation
  % w_\varsigma=
  %     \prod_{\varsigma'\neq\varsigma}
  %     \big(\smash{\sum_{\idxE|\varsigma_\idxE\neq\varsigma'}}\Var[\ncor^{\varsigma'}_\idxE]\big)
  %     % \varneq_{\varsigma'}
  % \Big/
  % \Big\{
  % \sum_{\varsigma'}\prod_{\varsigma''\neq\varsigma'}
  %     \big(\smash{\sum_{\idxE|\varsigma_\idxE\neq\varsigma''}}\Var[\ncor^{\varsigma''}_\idxE]\big)
  %      %\varneq_{\varsigma''}
  % \Big\}.
  %
  % Simpler version
  % w_\idxE=
  %     \prod_{\idxE'\neq\idxE}
  %     \big(\smash{\sum_{\idxE\neq\idxE'}}\Var[\ncor^{\idxE'}_\idxE]\big)
  % \Big/
  % \Big\{
  % \sum_{\idxE'}\prod_{\idxE''\neq\idxE'}
  %     \big(\smash{\sum_{\idxE\neq\idxE''}}\Var[\ncor^{\idxE''}_\idxE]\big)
  % \Big\}.
  w_\idxE = \prod_{\idxE'\neq\idxE} \boldsymbol\theta_{\idxE'}
  \Big/
    \Big(
     \sum_{\idxE'}\prod_{\idxE''\neq\idxE'}\boldsymbol\theta_{\idxE''}
   \Big),
\end{gather}
where $\Var[\ncor^{\idxE'}_\idxE]\equiv (\ncor^{\idxE'}_\idxE)^2$ is the expected variance of
the correction term which is defined similarly to Ref.~\cite{Brandbyge2002}.
The derivation of the expression for $w_\varsigma$ is in the SM. Additionally, 12 different
weighing schemes have been checked to infer whether they might provide a better estimate
of $\DN$. However, we have found that the argumentation in Ref.~\cite{Brandbyge2002}
provides the best physical interpretation and also the best weighing. It is  outside
the scope of this paper to document their differences, yet they are available for end
users.

If bound states are present in the system we weigh each equilibrium contribution
equally. An example of the ambiguity of selecting the proper weight of bound states can be
found in the SM.

\subsection{Inversion algorithms and performance}
\label{ssec:implementation}

\siesta\ uses localized basis-orbitals (LCAO) which inherently introduce sparse
Hamiltonian, density, and overlap matrices.  The sparsity of the density matrix means that
one only needs to compute the Green function for the appropriate non-zero
elements. Further, in the NEGF formalism this computation relies on the inversion of the
Hamiltonian and the overlap matrices (and self-energies). It is then beneficial to utilize
specialized algorithms that can deal with this \emph{selected inversion}.

Several inversion strategies have been explored
\cite{Amestoy2000,Amestoy2001,Brandbyge2002,Petersen2008,CMS2009,Li2008,Lin2011,Lin2011a,Jacquelin2014,Jacquelin2015,Hetmaniuk2013,Okuno2013,Feldman2014,Thorgilsson2014}
which all have their advantages and disadvantages. The MUMPS and SelInv methods are very
efficient for calculating a small subset of the inverse matrix (EGF), while for dense
parts of the matrix they are less effective (NEGF)
\cite{Amestoy2000,Amestoy2001,Lin2011,Lin2011a,Jacquelin2014,Jacquelin2015}.  \tsiesta\
originally implemented direct inversion using LAPACK \cite{LAPACK,Brandbyge2002}.
%Only recently, have the SelInv algorithm enabled complex asymmetric matrix inversions which
%may prove promising for NEGF calculations in future works \cite{CMS2009}.

In the following we will present 3 different inversion
algorithms \cite{Godfrin1991,Amestoy2000,Amestoy2001,Brandbyge2002,Hod2006,Petersen2008,Reuter2012}, 1) direct (LAPACK), 2) sparse (MUMPS) and 3) block-tri-diagonal (BTD). The methods are all
implemented in two variants, $\Gamma$-only ($\kk=0$) and arbitrary $\kk$-point, to limit memory usage where applicable.
In the following we omit the explicit $\kk$-dependence without loss of generality. For the non-equilibrium
part of the contour a triple product of a Green function block column is required to
calculate the spectral function in the non-zero elements of the density matrix. To
calculate the spectral function, the needed block columns are those where $\Scat_\idxE$ is
non-zero, \ie
\begin{equation}
  \label{eq:Gf:Spec}
  \Spec_\idxE(z)=\G(z) \Scat_\idxE(z) \G^\dagger(z)=
  \bgroup
  \def\arraystretch{0.}%  1 is
  \setlength\tabcolsep{0pt}
  \begin{tabular}{|@{}ccc@{}|}
    \hline
    \sq & \sq & \sq
    \\
    \sq & \sq & \sq
    \\
    \sq & \sq & \sq
    \\
    \hline
  \end{tabular}\,
  \begin{tabular}{|@{}ccc@{}|}
    \hline
    \sq &  &
    \\
    & \color{white} \sq &
    \\
    & & \color{white} \sq
    \\
    \hline
  \end{tabular}\,
  \begin{tabular}{|@{}ccc@{}|}
    \hline
    \sq & \sq & \sq
    \\
    \sq & \sq & \sq
    \\
    \sq & \sq & \sq
    \\
    \hline
  \end{tabular}^\dagger
  =
  \begin{tabular}{|@{}ccc@{}|}
    \hline
    \sq &  &
    \\
    \sq & \color{white} \sq &
    \\
    \sq &  & \color{white} \sq 
    \\
    \hline
  \end{tabular}\,
  \begin{tabular}{|@{}ccc@{}|}
    \hline
    \sq &  &
    \\
    & \color{white} \sq &
    \\
    & & \color{white} \sq
    \\
    \hline
  \end{tabular}\,
  \begin{tabular}{|@{}ccc@{}|}
    \hline
    \sq &  &
    \\
    \sq & \color{white} \sq &
    \\
    \sq &  & \color{white} \sq 
    \\
    \hline
  \end{tabular}^\dagger
  \egroup.
\end{equation}
Equation~\eqref{eq:Gf:Spec} is implemented for the 3 inversion algorithms which substantially
reduces memory requirements, and particularly so for the BTD method.
% %
% The LAPACK method is easily implemented while the MUMPS method requires a specific data
% layout of the sparse matrix format. MUMPS is also used in other software for Green
% function techniques \cite{Groth2014,ATK}.

\subsubsection{Block-tri-diagonal inversion}
\label{ssec:BTD}

Our block-tri-diagonal matrix inversion algorithm has become the default method in
\tsiesta. This algorithm was originally described in Ref.~\cite{Godfrin1991} while we
follow the simpler outlined form in Refs.~\cite{Hod2006,Reuter2012}. Often, this method is
known as the recursive Green function method which corresponds to creating a quasi 1D,
block tri-diagonal matrix.

\begin{figure}
  \centering
  \includegraphics{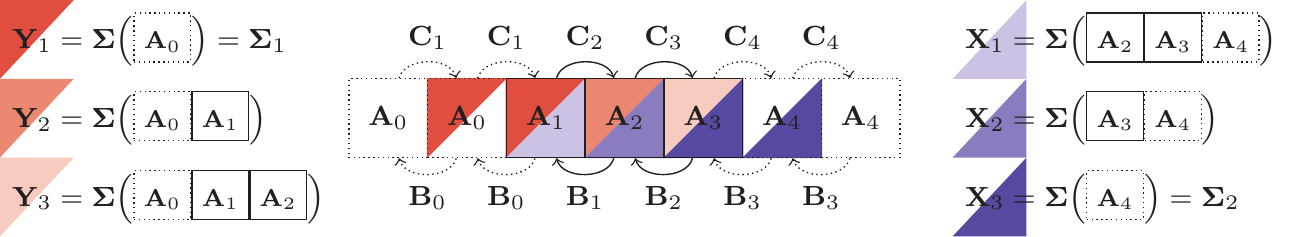}
  \caption{Block-tri-diagonal inversion algorithm shown in terms of the Hamiltonian
      elements of a 2-electrode system. 
      The inverse Greens function consists of block matrices in the diagonal and lower/upper diagonals denoted by $\mathbf A_i$ and $\mathbf B_i$/$\mathbf C_i$, respectively. The surface self-energies are calculated from the bulk electrode and the self-energies are propagated through the system, allowing one to calculate the exact Green function in any block of the infinite
      matrix. 
      \label{fig:BTD-inv}
  }
\end{figure}

The algorithm can be illustrated as shown in \fref{fig:BTD-inv} and follow these equations
\begin{align}
  \label{eq:BTD:algo}
  \mathbf \G^{-1} &=
  \begin{pmatrix}
    \mathbf A_1 & \mathbf C_2 & 0 & \cdots & \\
    \mathbf B_1 & \mathbf A_2 & \mathbf C_3 & 0& \cdots \\
    0 & \mathbf B_2 &\ddots & \ddots & 0 \\
    \vdots & 0  & \ddots &  \ddots & \mathbf C_p \\
     & \vdots  & 0  & \mathbf B_{p-1} & \mathbf A_p\\
  \end{pmatrix}, &\quad
  \begin{aligned}
    \widetilde{\mathbf Y}_n&=\left[\mathbf A_{n-1}-\mathbf Y_{n-1}\right]^{-1}
    \mathbf C_n, %
    & & \mathbf Y_1=0,
    \\
    \mathbf Y_n&=\mathbf B_{n-1}\widetilde{\mathbf Y}_n.
    \\
    \widetilde{\mathbf X}_n & =\left[\mathbf A_{n+1}-\mathbf X_{n+1}\right]^{-1}
    \mathbf B_n, %
    & & \mathbf X_p=0,
    \\
    \mathbf X_n & =\mathbf C_{n+1}\widetilde{\mathbf X}_n,
  \end{aligned}
\end{align}
$\mathbf A_i$, $\mathbf B_i$ and $\mathbf C_i$ correspond to the non-zero elements of
$z\SO - \HH - \sum_\idxE\SE_\idxE$, cf.~\eref{eq:Gf}. $\mathbf Y_n$/$\mathbf X_n$ can be
thought of as the self-energies connecting to a previous sequence of
$\mathbf Y_m$/$\mathbf X_m$ for $n>m$/$m>n$. These are sometimes denoted the ``downfolded''
self-energies and are indicated in \fref{fig:BTD-inv}. For instance $\mathbf Y_2$
corresponds to a self-energy of an infinite bulk part connecting to $\mathbf A_1$. Note
that only for a Left/Right terminal system with strict ordering of orbitals will
$\mathbf A_1=z\SO_{1,1} - \HH_{1,1} - \SE_{\mathrm{Left}}$ and
$\mathbf A_p=z\SO_{p,p} - \HH_{p,p} - \SE_{\mathrm{Right}}$. Importantly the
$\widetilde{\mathbf Y}_i$/$\widetilde{\mathbf X}_i$ matrices can be calculated using a
linear solution instead of an inversion and subsequent matrix-multiplication\footnote{One
    may solve a set of linear equations:
    $\protect\big[\protect\mathbf A_{n-1}-\protect\mathbf Y_{n-1}\protect\big]\protect\widetilde{\protect\mathbf Y}_n = \protect\mathbf C_n$ and
    $\protect\big[\protect\mathbf A_{n+1}-\protect\mathbf X_{n+1}\protect\big]\protect\widetilde{\protect\mathbf X}_n = \protect\mathbf B_n$.}.
The strict ordering of the self-energies is not a requirement and $\SE_\idxE$ may be split
among any sub-matrices\footnote{$\protect\SE_\protect\idxE$ can maximally be split in two consecutive
    blocks as it is a dense matrix.} $\mathbf A_i$, $\mathbf B_i$ or $\mathbf C_i$.
Calculating any part of the Green function then follows the iterative solution of these
equations
\begin{align}
  \label{eq:BTDs}
  \left.
    \begin{aligned}
      \G_{n,n}&=\left[\mathbf A_n-\mathbf X_n-\mathbf Y_n\right]^{-1},
      \\
      \G_{m-1,n}&%=-\left[\mathbf A_m-\mathbf Y_m\right]^{-1}\mathbf C_{m+1} \G_{m+1,n}
      =
      -\widetilde{\mathbf Y}_{m} \G_{m,n}\quad\text{ for }m\le n,
      \\
      \G_{m+1,n}&%=-\left[\mathbf A_m-\mathbf X_m\right]^{-1}\mathbf B_{m-1} \G_{m-1,n}
      =
      -\widetilde{\mathbf X}_{m} \G_{m,n}\quad\text{ for }m\ge n,
    \end{aligned}\qquad
  \right\}\quad\G = 
  \begin{aligned}
    \includegraphics{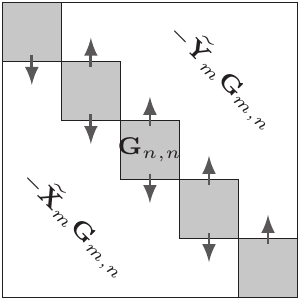}
  \end{aligned}
\end{align}
The algorithm for the above calculation is shown in \fref{alg:Gf}.

% \aref{alg:Gf} describes the algorithms implemented for the BTD method. For the equilibrium
% density we only need the Green function calculation (left). This is straightforward with a
% linear progress.

For the non-equilibrium part the straightforward implementation of the column product in
\eref{eq:Gf:Spec} involves calculating the full Green function column for columns of the
scattering matrix and a subsequent triple matrix product for each block. However, it may
be advantageous to utilize the propagation of the spectral function by using
Eqs.~\eqref{eq:BTDs} which inserted into \eref{eq:Gf:Spec} yields
\begin{equation}
  \label{eq:BTD:A-propagate}
  \Spec_{\idxE} =
  \begin{aligned}
    \includegraphics{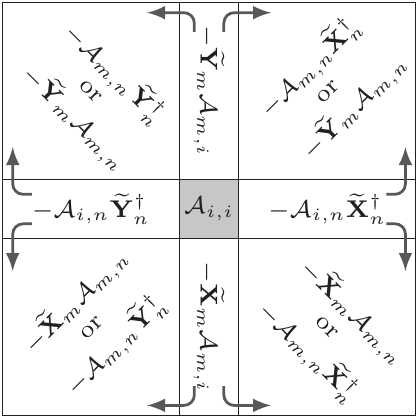}
  \end{aligned}.
\end{equation}
Recall that $\Spec_{\idxE,i,i}=\G_{i,i}\Scat_\idxE\G^\dagger_{i,i}$.
Importantly Eqs.~\eqref{eq:BTD:A-propagate} are recursive equations similar to
Eqs.~\eqref{eq:BTDs}. The propagation of the spectral function is often faster than the
straightforward method. The algorithm for the propagation method is shown in
\fref{alg:Gf}.  To reduce computations we calculate $\wX_i$, $\wY_i$ and the diagonal
Green function elements where $\Scat_\idxE$ lives for all $N_\idxE$ and store these
quantities in one BTD matrix. Subsequently we calculate the spectral function for each
electrode separately in another BTD matrix (without re-calculating $\wX_i$, $\wY_i$) to
drastically reduce computations.

It is important to note that the required elements of the density matrix are only those of
the block-tri-diagonal matrix ($\G$ and $\Spec_\idxE$) corresponding to the elements shown
in \eref{eq:BTD:algo}. For $\G$, \eref{eq:BTDs}, one only calculates $\G_{i,i}$,
$\G_{i+1,i}$ and $\G_{i,i+1}$ and similarly for $\Spec_\idxE$. For the latter we only use
the upper-left and lower-right algorithms as presented in \eref{eq:BTD:A-propagate}.

\begin{figure}
  \centering
  \newcommand\A{\mathbf A}
  \newcommand\B{\mathbf B}
  \newcommand\C{\mathbf C}
  \newcommand\M{\mathbf M}
  \newcommand\N{\mathbf N}
  \newcommand\X{\mathbf X}
  \newcommand\Y{\mathbf Y}

  \def\myheader#1{\centering \underline{#1}\vspace{1ex}}
  \algnewcommand\algorithmictrue{\textbf{True}}
  \algnewcommand\algorithmicfalse{\textbf{False}}
  \begin{minipage}[c]{0.375\linewidth}
    \myheader{EGF}
    \begin{algorithmic}[1]
      \State allocate $\M$ and $\N$ with $p$ blocks
      \ForAll{$\varsigma\in N_\varsigma$}
      \State initialize all $\A_i$, $\B_i$, $\C_i$ in $\M$
      \ForAll{$i\in\{1,\dots,p-1\}$}
      \State $\N_{i+1,i}\gets\widetilde\Y_{i+1}$
      \State $\N_{p-i,p-i+1}\gets\widetilde\X_{p-i}$
      \EndFor
      \ForAll{$i\in\{1,\dots,p\}$}
      \State $\N_{i,i}\gets\G_{i,i}$
      \EndFor
      \ForAll{$i\in\{1,\dots,p-1\}$}
      \State $\N_{i+1,i}\gets\G_{i+1,i}$
      \State $\N_{p-i,p-i+1}\gets\G_{p-i,p-i+1}$
      \EndFor
      \State $\DE^\varsigma \gets \N$
      \EndFor
    \end{algorithmic}
  \end{minipage}\qquad
  \begin{minipage}[c]{0.55\linewidth}
    \myheader{NEGF}
    \begin{algorithmic}[1]
      \State allocate $\M$ with $p$ blocks
      \State allocate $\N$ with $p$ blocks
      \State initialize all $\A_i$, $\B_i$, $\C_i$ in $\M$
      \ForAll{$i\in\{1,\dots,p\}$}
      \ForAll{$\idxE\in\{1,\dots,N_\idxE\}$}
      \If{$\mathrm{any}(\mathrm{column}(\idxE))\in \M_{i,i}$}
        \State $\mathrm{calc}(i) \gets \algorithmictrue$
      \Else
        \State $\mathrm{calc}(i) \gets \algorithmicfalse$
      \EndIf
      \EndFor\EndFor
      \ForAll{$i\in\{1,\dots,p-1\}$}
      \State $\N_{i+1,i}\gets\widetilde\Y_{i+1}$
      \State $\N_{p-i,p-i+1}\gets\widetilde\X_{p-i}$
      \EndFor
      \ForAll{$i\in\mathrm{calc}(:)=\algorithmictrue$}
      \State $\N_{i,i}\gets\G_{i,i}$ \Comment{Only columns needed}
      \EndFor
      \ForAll{$\idxE\in \{1,\dots,N_\idxE\}$}
      \State $\M_{i,i}\gets \G_{i,i}\Scat_\idxE\G_{i,i}^\dagger$
      \ForAll{$\{m,n\}\in\{1,\dots,p\}$}
      \If{$m+n<2i$}
      \State $\M_{m,n} \gets
      \wY_{m+1}\cdots\wY_i\G_{i,i}
      \Scat_\idxE
      \G_{i,i}^\dagger
      \wY_i^\dagger\cdots \wY_{n+1}^\dagger$
      \Else
      \State $\M_{m,n} \gets
      \wX_{m-1}\cdots\wX_i\G_{i,i}
      \Scat_\idxE
      \G_{i,i}^\dagger
      \wX_i^\dagger\cdots \wX_{n-1}^\dagger$
      \EndIf
      \State $\M_{m,n} \gets \M_{m,n}(-1)^{m+n}$
      \EndFor
      \State $\DN^\idxE \gets \M$
      \EndFor
    \end{algorithmic}
  \end{minipage}
  \caption{The algorithm used for inverting a generic BTD matrix as well as columns for
      calculating the spectral function in NEGF calculations. The EGF algorithm is a
      direct recursive algorithm \cite{Reuter2012}, while the NEGF algorithm is a
      modification to reduce the memory requirement for calculating the spectral function.
      \label{alg:Gf}
  }
\end{figure}

\subsubsection{Orbital pivoting for minimizing bandwidth}
\label{ssec:BTD:pivot}

The performance of the BTD algorithm is determined solely by the bandwidth of the
Hamiltonian matrix, \ie\ the size of the $\mathbf A_n$ blocks. The bandwidth is an
expression of the quasi 1D size of the system and is defined as
\begin{equation}
  \label{eq:bandwidth}
  B(\mathbf M) = \max\big(|i - j| \big|\mathbf M_{ij}\neq0\big).
\end{equation}
Internally, the sparsity pattern in \siesta\ is determined via the atomic input
sequence. However, this sparsity pattern will rarely have the minimum bandwidth,
particularly so for $N_\idxE\neq2$. To minimize the matrix bandwidth, and increase
performance, we have implemented 5 different pivoting methods, 1) connectivity graph based
on the Hamiltonian sparse pattern, 2) peripheral connectivity graph based on a
longest-path solution before a connectivity graph between end-points, 3) Cuthill-Mckee
\cite{Cuthill1969}, 4) Gibbs-Poole-Stockmeyer \cite{Gibbs1976}, and 5) generalized
Gibbs-Poole-Stockmeyer \cite{Wang2009}. The first 2 are developed by the authors and
exhibit a good bandwidth reduction of the matrix for a majority of systems. The latter 3
methods are used in interaction graphs with few nodal points which may be the reason for
their, sometimes, poor bandwidth reduction capability in atomic structure
calculations. For cases of many nodal points per point, such as for 3D bulk structures, each of the methods yield
different optimal orderings. Currently there exists no omnipotent method for bandwidth
reduction and we encourage checking the different methods for each system. One may greatly
increase pivoting performance by using the atomic graph, rather than the orbital
graph. Indeed there is, obviously, little to no difference between the atomic and orbital
graphs.

\begin{figure}
  \centering
  {\footnotesize
      \begin{tabular}[c]{c|c|c}
        a)\hfill$N_\idxE=2$\hfill\phantom{}
        &
        b)\hfill$N_\idxE=3$\hfill\phantom{}
        &
        c)\hfill$N_\idxE=4$\hfill\phantom{}
        \\
        \hfill
        \includegraphics{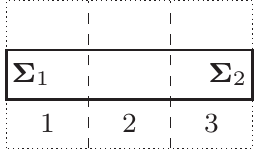}\hfill\phantom{}
        &
        \hfill
        \includegraphics{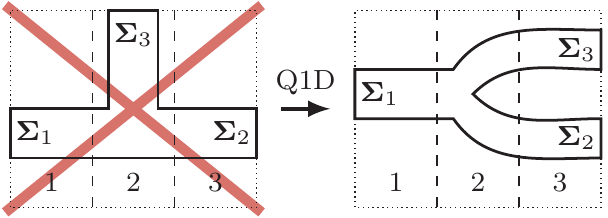}\hfill\phantom{}
        &
        \hfill
        \includegraphics{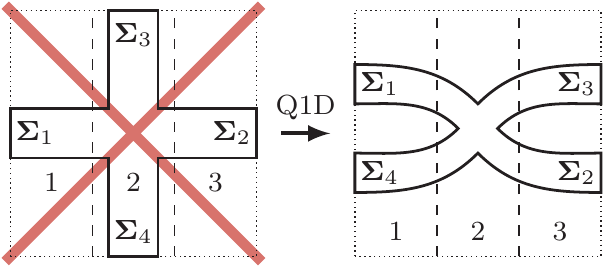}\hfill\phantom{}
      \end{tabular}
  }
  \caption{Quasi 1D partitioning in 3 parts (divided by dashed lines), for varying number
      of electrodes. The dotted lines denote the cell boundary and the fully drawn lines
      encompass the atoms/Hamiltonian elements. $\SE_i$ are the self-energies that couple
      the device to the semi-infinite electrodes. The crossed illustrations are the naive
      partitioning of the quasi-1D system (the naive partitioning for $N_\idxE=2$
      \emph{is} the best partitioning), whereas a better quasi-1D pivoting is also
      shown. Note that in \emph{any} of the systems shown, the ordering of the
      self-energies can be swapped \emph{at will}. 
      \label{fig:quasi-1D}
  }
\end{figure}

Pivoting becomes increasingly important when considering $N_\idxE>2$ electrodes as the
quasi 1D block-partitioning becomes less obvious. In \fref{fig:quasi-1D} we illustrate the
naive block partition for $N_\idxE=\{2,3,4\}$ together with an improved partitioning. For
$N_\idxE=2$ the naive \emph{is} a good partitioning. The naive $N_\idxE=3$ problem will
create a big block for $p=2$ which will decrease performance. However, by grouping two
electrodes the quasi-1D problem can be much improved. Grouping should be chosen to
minimise all block sizes, \emph{e.g.} if the self-energy bandwidth, \eref{eq:bandwidth},
of $B(\SE_2)>B(\SE_1)+B(\SE_3)$ then branches $\SE_1$ and $\SE_2$ should swap places in
\fref{fig:quasi-1D}b. Similarly for $N_\idxE=4$ two groups occur for both ends of the
quasi 1D matrix. The grouping of electrodes can easily be generalised for any $N_\idxE$
electrodes dependent on the branch sizes.

Pivoting complicates the triple product \eref{eq:Gf:Spec} due to partitioning of the
scattering matrix with respect to the Green function. However, each block $\mathbf A_n$
can be sorted such that $\Scat_\idxE$ becomes consecutive in memory for optimal
performance. This will maximally split $\Scat_\idxE$ into two blocks. We stress that
splitting the broadening matrices, $\Scat_\idxE$, up into two blocks, if possible, is more
beneficial than retaining a single block because the bandwidth of the tri-diagonal matrix
will be smaller.

\subsubsection{Performance of inversion algorithms}
A timing comparison of the three methods against \tsiesta\ 3.2 \cite{Brandbyge2002} is shown in
\fref{fig:perf}. A pristine graphene system is used with square electrodes of 48 atoms,
corresponding to $2\times 6$ (zigzag by armchair directions, respectively). Figure \ref{fig:perf}a show the
timing for differing lengths of pristine graphene up to $\sim$6,000 orbitals\footnote{A
    comparison for larger systems is not possible due to \protect\tsiesta\ 3.2 \protect\cite{Brandbyge2002} memory
    consumption.} using an EGF calculation. Figure \ref{fig:perf}b is the same calculation but with a high applied
bias of $\unitr{0.75}V$ (resulting in an equal amount of non-equilibrium and equilibrium contour points).
%
% \change{All three methods are \emph{much} faster than the previous version}{there is a
%     certain degree of Transiesta bashing in the manuscript that I like very much, but I am
%     not sure other authors would appreciate: do we need to say that the previous version
%     was not very efficient?}\change{}{No not really, it was never intended as bashing,
%     more as improving statements. However, as we retain the \tsiesta\ name my idea was
%     that that was a great acknowledgement of the previous work? I am of course open to
%     suggestions}
%
% The LAPACK version is seen to be $\sim10$ times faster for NEGF while
% just slightly faster in the EGF part. 
MUMPS performs very well for EGF while NEGF
is rather inefficient due to clustering of columns. Further studies of MUMPS have shown
that it performs better for more than $\sim$5,000 orbitals as the
sparsity increases\footnote{The MUMPS comparison is thus not justifying the actual
    performance for large systems.}. Lastly, the BTD method is performing extremely well
reaching around 100 times better performance on systems at 5,000 orbitals.  Doing even
larger systems will only increase the speedup even more so. In all investigated systems we
have found an impressive speedup for the BTD method.
%
%The current implementation has also been compared against commercial software and
%performance is at least on par with ATK\cite{ATK}.

\subsubsection{Parallelization}

\begin{figure}
  \centering
  \includegraphics{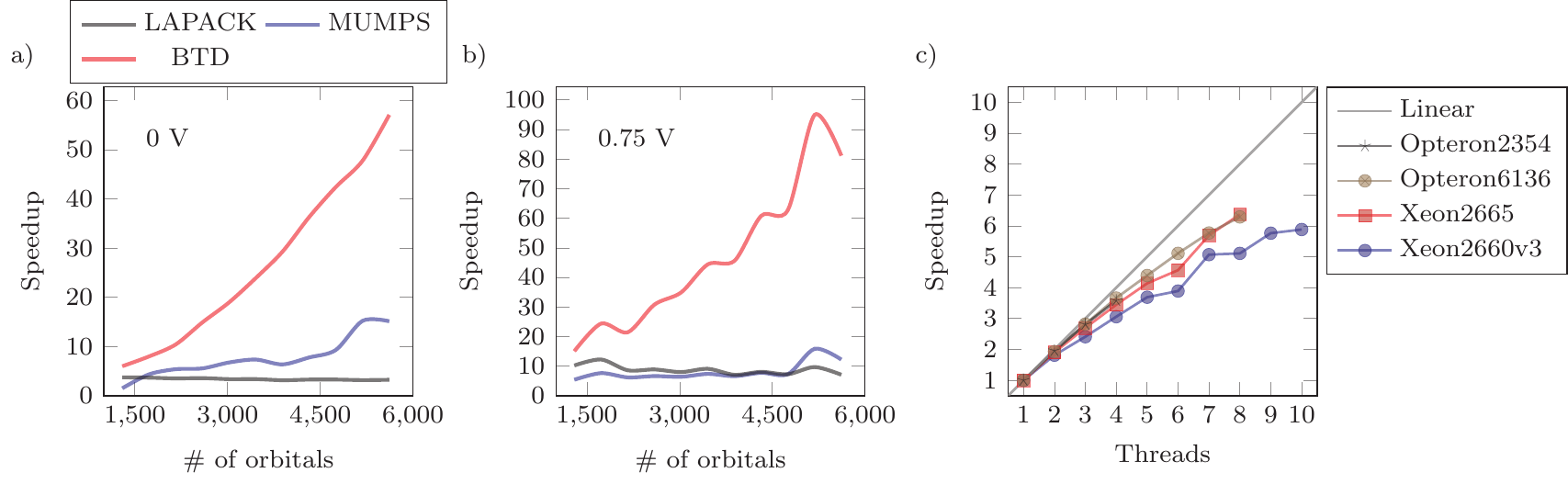}
  \caption{Performance characterization of \tsiesta\ using a pristine graphene
      cell (24 atoms wide). Speedup for (a) EGF and (b) NEGF calculations of pristine graphene 
      compared against the direct LAPACK implementation. The BTD method exhibits more than
      40 times the speed of the LAPACK implementation for the largest size. MUMPS gains
      speed after $5,000$ orbitals.
      (c) Threading performance using different hardware architectures
      running on a single node. Our test system has roughly 830 orbitals per BTD block and consists of 11 blocks in total. As the threading performance primarily stems from the threaded BLAS library
      one can see that the threading reaches a limit due to the rather small
      blocks. 
      \label{fig:perf}
  }
\end{figure}

Our inversion methods are parallelized across energy points, meaning that each MPI process
handles one energy point on the contour, but needs to hold the complete (non distributed)
matrices in memory. As the matrices dealt with in \tsiesta\ can become of GB size
depending on the width of the electrodes, this parallelization scheme might hit the
physical memory limit. To circumvent this we have updated the \tsiesta\ code to enable full
hybrid parallelization with OpenMP 3.1 threading\footnote{In \siesta\ threading has only
    been implemented in a few places, with priority on grid operations.}. Thus instead of
using $N_{\mathrm{tot}}$ MPI processes and reaching the memory limit, one can use
$N_{\mathrm{tot}}/N_{\mathrm T}$ processes and $N_{\mathrm T}$ threads per process. The
threads pool their associated memory resources, pushing the practical limit by a factor of
$N_{\mathrm T}$, and linear scaling is retained.
Figure \ref{fig:perf}c shows the threading performance for \tsiesta\ on different
hardware\footnote{We have used OpenBLAS 0.2.15 with OpenMP threading using the GNU 5.2.0
    compiler for a graphene test system. We use a non-threaded LAPACK. High compiler optimizations are used.} using a
single node, hence MPI-communication can be considered negligible for hardware with
multiple sockets. In our implementation an increase in the number of MPI processes would not affect the
threading performance, however additional MPI processes would increase MPI communication
time, thus favoring threading for large number of MPI processes. 

The graphene test system consists of 11 BTD blocks with an average block size of $N_B=830$ using the $\kk$-space version of the BTD method with a bias. We expect this system to represent a typical medium sized system of 9,130 orbitals. We find that there is extremely good scaling up to 4
threads. For higher number of threads the scaling is still good, but diverges. Threading is optimal
for $N_B/N_{\mathrm T}\ggg1$ which is one reason for the limiting speedup for large
thread-counts in the shown data. By calculating the parallel fraction using Amdahl's law
we get roughly 95\% across the investigated range of $N_{\mathrm T}$.

% We note that advanced CPU extensions (vectorization) of
% the Intel processors further leverages the threading performance for 2660v3. Only for even
% wider problem sizes will the threading limitation be relieved for high performance
% processors. This aptly demonstrates when hybrid parallelization can be of greater benefit
% and when one should revert to pure MPI parallelism. 

% Future improvements of the \tsiesta\ implementation would be to parallelize the MUMPS
% and BTD methods for increased versatility. This has been outside the scope of this work.

% \subsection{The so-called transport direction}
% \label{ssec:transport:direction}
% \input{transport-dir.tex}

\subsection{Bloch theorem and self-energies}
\label{ssec:SE}
A performance-critical part of Green function implementations is an efficient calculation
of the electrode self-energies. It is beneficial both computationally and in memory cost to calculate the self-energy using the smallest unit-cell which can be repeated to form the larger super-cell corresponding to the electrode-device contact region.
In \tsiesta\ we utilize Bloch's theorem and the corresponding $\kk$-point sampling when transverse periodicity is present.  The
Bloch expansion may be used for both the electrode Hamiltonian \emph{and} the
self-energies, given that the electronic structure is calculated at equivalent $\kk$
sampling, \emph{i.e.}, that an electrode which repeats out $X\times Y$ times in the larger device
    unit-cell, must be computed on $\kk$ mesh which is $X\times Y$ more dense.
Bloch expansion along one cell vector may be written in this short matrix form
\begin{equation}
  \label{eq:SE:bloch-expand}
  \SE_{k_n}^n =\frac1n
  \;
  \sum_{
      \mathclap{
          \substack{j\\
              k_j=k_n+2\pi\frac{j-1}{nR}
          }
      }
  }^n
  \qquad
  \begin{bmatrix}
    1
    &
    e^{-i k_jR}
    &
     \cdots
    &
    e^{-i nk_jR}
    \\
    e^{i k_jR}
    &
    1
    &
     \cdots
    &
      e^{-i (n-1)k_jR}
    \\
    \vdots
    &
     \vdots
    &
     \ddots
    &
     \vdots
    \\
     e^{i nk_jR}
    &
     e^{i (n-1)k_jR}
    &
    \cdots
    &1
  \end{bmatrix}
  \otimes
  \SE_{k_j}^1,
\end{equation}
where $\SE^n$/$\SE^1$ is the self-energy in the larger/smallest unit-cell and $n$ is the
number of times the smallest unit-cell is repeated to coincide with the larger
unit-cell. Here $R$ denotes the cell length of the small unit-cell. Lastly, $k_n$ is the
$k$-point in the repeated super-cell.
From \eref{eq:SE:bloch-expand} it can be inferred that one needs to calculate the
self-energy $\SE^1$ for $n$ $\kk$ points instead of calculating $\SE^n$ once. However,
calculating the self-energy scales cubic and using a smaller matrix is far more
beneficial.

\subsection{Electrostatics in NEGF}
\label{ssec:Hartree}
The Hartree electrostatic potential plays an essential role for the NEGF calculations.  In
\tsiesta\ it is determined from the difference between the self-consistent
electron density $\rho(\rr)$ and the neutral atom density $\rho^{\mathrm{atom}}(\rr)$ and
solved using a Fourier transformation of the Poisson equation.
% For charge neutral systems
% ($\int\drho(\rr)=0$) this generates a single contribution to the electrostatic potential
% $\nabla^2\dV_0(\rr)=-\drho(\rr)/\epsilon_0$. For charged systems the potential is obtained
% by adding a constant term which corresponds to a uniform charge background that
% compensates the excess charge, \ie\ $\dV(\rr)=\dV_0(\rr)+\dV_{\mathrm{back}}$. 
%The calculation cell is periodic in all directions and thus slab calculations require a
%vacuum to electrostatically decouple its periodic replica and using a slab dipole
%correction is a requirement\cite{Bengtsson1999}. As \siesta\ uses localized basis
%orbitals vacuum is inexpensive, contrary to plane wave codes.
%
\tsiesta\ implements a generic interface to correctly introduce the appropriate boundary
conditions, fully controlled by the user. 
%A generic
%Poisson solver with arbitrary boundary conditions inherently performed in \tsiesta\ is
%thus outside the scope of this work.
% For \tsiesta\ this introduces erroneous boundary conditions and other solvers may be added
% in the future to solve this deficiency. Such solvers are beyond the scope of this
% work. \comment{AG: As it stands, this comment sounds really bad. It looks as if \tsiesta\ cannot give
% good results due to a fundamental problem...}

For a reasonable description of the electrostatics in the NEGF setup one generally
requires that the electrode regions in the device essentially behave as bulk. This
requirement ensures a smooth electrostatic interface between the device and the
semi-infinite, enforced bulk electrodes. For metallic electrodes this may easily be
accomplished as the electronic screening length is short (typically a few atomic
layers). Additionally we allow buffer atoms which are non-participating atoms in the
scattering region calculation. They may be used to screen electrodes such that smaller
scattering regions may be used. Such constructs are useful when non-periodic or dissimilar
electrodes are used.
%
%\revision{If non-periodic electrodes are used, (isn't this something else?
%    buffers are also useful in combination with the Bloch expansion?)} 
% How are they more useful for Bloch expansion?
% I have only used them in two cases:
% 1. when one wishes to reduce the scattering region size and
% ABCABC-molecule-BCABCA is used. Then one requires BC as buffer atoms. 
% 2. When dissimilar electrodes are used.
Along side with buffer atoms several other methodologies for improving the
electrode/device interface exists, such as forcing the density to be bulk, or calculating
$\DM$ in the electrode region.
%
% Generally the system electrode region has a $\rho(\rr)$ very similar to that of a bulk
% calculation if the electrode is metallic and is an effective screening medium. For systems
% with low screening and/or using buffer atoms as outlined in \sref{ssec:buffer} $\rho(\rr)$
% might not behave as \emph{bulk} like as wanted. When this is the case the calculated
% Hartree potential can be inferior represented to that expected. \tsiesta\ allows one to
% force $\rho(\rr)$ in electrode regions to be that of the electrode bulk density
% calculation. This will better reproduce the potential profile interface between the device
% and the bulk electrode.

\begin{figure}
  \centering
  \includegraphics{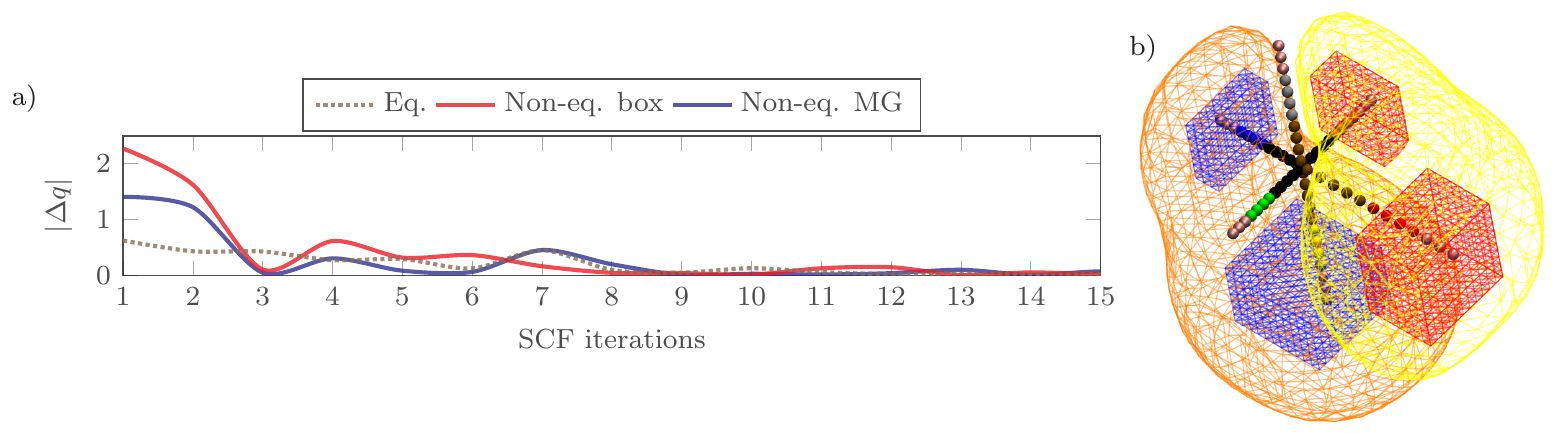}
  \caption{a) Charge conservation with respect to \# of SCF iterations for a $N_\idxE=6$
      device (b). Both zero bias and two different initial guesses of the Hartree
      potential $V_H(\rr)$. The dashed curve is for equilibrium while the full lines are
      an electrode box guess and a full MG solution. Providing a better guess improves
      convergence. b) also shows the initial MG guess for 4 iso-values $\pm V/2$ and
      $\pm V/10$.
      \label{fig:charge}
  }
\end{figure}

In open boundary calculations, such as NEGF, one also needs to ensure that the Hartree
potential fulfills the specific boundary conditions at the electrodes. We employ a
formulation similar to the original implementation \cite{Brandbyge2002,Rocha2005}. For
$N_\idxE=2$ and a shared semi-infinite direction a linear potential ramp can be used as a
guess \cite{Brandbyge2002}.
%  can easily be described using the following partition with $x$
% being the transport direction
% \begin{equation}
%   \label{eq:Vh:2}
%   V_H(x)=\tilde\phi(x)+
%   \begin{cases}
%     \mu_L &\quad\text{, for $x<x_{LC}$}
%     \\
%     \mu_R &\quad\text{, for $x_{CR}<x$}
%     \\
%     (\mu_R-\mu_L)\frac{x-x_{LC}}{x_{CR}-x_{LC}} + \mu_L &\quad\text{, else,}
%   \end{cases}
% \end{equation}
% where $x_{LC}$, $x_{CR}$ are the transport direction coordinates that intersect a plane
% between the left electrode and the device region, and the right electrode and the device
% region, respectively. Equation~\eqref{eq:Vh:2} assumes the electrodes to be bulk, whereas the same
% equation for non-bulk electrodes would be simpler,
% \begin{equation}
%   \label{eq:Vh:2:non-bulk}
%   V_H(x)=\tilde\phi(x)+ V\Big(\frac12-\frac{x}{L_x}\Big).
% \end{equation}
%
%
For unaligned semi-infinite directions the boundary conditions become non-trivial. The
simplest initial guess for the Hartree potential to fulfill the boundary conditions is a
\emph{box} guess
\begin{equation}
  \label{eq:Vh:N}
  V_H(\rr)\leftarrow V_H(\rr)+\sum_\idxE
  \begin{cases}
    \mu_\idxE &\quad\text{, for $\rr\in\rr_\idxE$}
    \\
    0 &\quad\text{, for $\rr\not\in\rr_\idxE$}
  \end{cases}
\end{equation}
where $\rr_\idxE$ denotes the part in the real space grid where the electrode atoms
reside. However, this introduces non-smooth potentials between the electrode and device
region and it may only yield qualitative approximations to the actual Hartree potential.
% The favor
% of using \eref{eq:Vh:N} is its simplicity contrary to solving the Poisson equation with
% arbitrary boundary conditions.
%
%
\tsiesta\ also allows custom Hartree potentials for improving convergence.
It is recommended to provide a custom guess for $N_\idxE>2$ as charge conservation and
convergence can easily be improved. Note that the initial guess for the Hartree solution
is linear in the difference between $\mu_\idxE$, hence only one guess calculation of the
Hartree potential is needed which makes this an in-expensive setup calculation.
% TF: not sure I understand
% NRP: Say you have a system: elec-1 = -1/2 V, elec-2 = 0V, elec-3 = 1/2V. 
% You may then calculate the potential profile at 1V [V_H(1V)]
% The solution at 0.5 V [V_H(0.5V)] will be:
%   V_H(0.5V) = V_H(1V) / 1V * 0.5V
% Thus you ever only need to calculate one.
% This will work if the chemical potentials are linearly dependent on $V$.
% However, if you have one lead with a fixed chemical potential, the linearity is not applicable.
% 
We have implemented a multigrid (MG) solver for the Poisson equation and applied it to a
$N_\idxE=6$ system with three different chemical potentials: [$-V/2$, $0$, $V/2$], see
\fref{fig:charge}b. The grid overlaying the geometry corresponds to the guessed Poisson
solution for the MG method at 4 iso-values, $V/2$ (blue), $-V/2$ (red), $V/10$ (orange)
and $-V/10$ (yellow). The Poisson solution is linearly dependent on $V$ due to the linearity of the
chemical potentials. The heavily colored atoms are electrodes, ``behind'' each electrode
are 3 buffer atoms to retain a bulk-like electrode. It consists of three crossing linear
chains with one carbon chain, one gold chain and one half-half carbon-gold chain.
In \fref{fig:charge}a we plot the absolute charge difference from the expected charge
after each SCF iteration for both the equilibrium case and for an applied bias of \unit1V
using two different initial guesses. All calculations settles after 8 iterations at nearly
no charge difference. For the two non-equilibrium cases the custom MG method reduces the
initial fluctuations compared to the \emph{box} guess. 

% Finally we remark that the potential profile for $N_\idxE$ calculations need thorough
% thought and rigorous testing of the convergence. A further investigation of the Hartree
% potential and its best solution lies outside the scope of this work.

In addition to the electrostatics associated with complex boundary conditions we also
extend \siesta\ (and \tsiesta) by enabling electrostatic gates, as described in
Ref.~\cite{Papior2016a}. This allows to introduce additional non-interacting electrodes to act as gates.
%
% As an extension for \siesta\ we have implemented 2 different gating methods which can also
% be used with the NEGF scheme.
%
Several implementations of electrostatic Hartree gates use an explicit
Hartree term $V_H(\rr)$ \cite{ATK,Ozaki2010} while other implementations deal with the
additional electrostatic terms arising from the charge distribution in the gate material
\cite{Otani2006}.
We have implemented both a Hartree gate and a charge gate with few restrictions on the geometry of
the gate. The geometries includes spherical, planes, rectangles and/or boxes are all allowed, further details can
be found in Ref.~\cite{Papior2016a}. The Hartree gate is similar to that in Refs.~\cite{ATK,Ozaki2010}
while the charge gate is a phenomenological model resembling Refs.~\cite{Otani2006,Brumme2014}.
%
%The Hartree gate is straightforwardly implemented and can be used to screen electrons in
%regions of high/low potential. 
%
Due to the simplicity of the Hartree gate we will instead focus here on explaining the charge gate
method, with which one can simulate complex gate configurations in both NEGF
and regular, 3D-periodic DFT calculations \cite{Papior2016a}.

\begin{figure}
  \centering
  \includegraphics{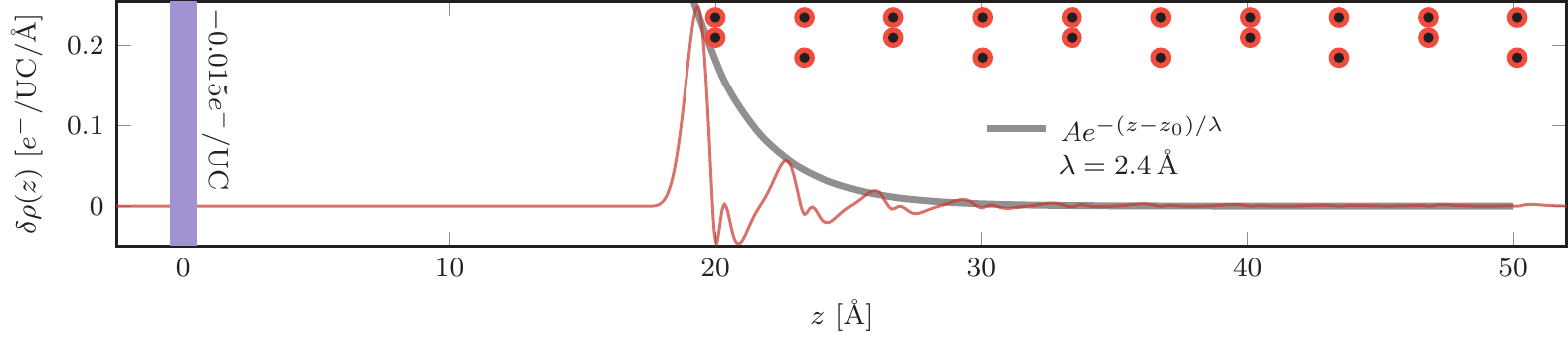}
  \caption{Electronic density decay length of a $10$-layer graphene stack. The red-black circles shows the placement of the carbon atoms
      (A-B stacking). The thin line show the difference in electronic density between the
      gated and non-gated system. The
      thick line shows a fitted decay profile of the gate-induced electronic density. A
      decay length of $2.4\,\text\AA$ is found at this particular gating level. 
      \label{fig:gating:graphite}
  }
\end{figure}

Figure~\ref{fig:gating:graphite} shows an example of a charge gate introduced in DFT
periodic calculations containing $10$ graphene layers (AB stacking). 
The charge gate is placed $20\,\text\AA$ away from (and parallel with) the first graphene
layer. A strong gate corresponding to $0.015e^-$ per graphene unit-cell is applied and the
resulting charge re-distribution is dependent on the screening length of graphite. By fitting an
exponential function to the charge response we calculate the decay length for the
electronic screening length of the electric field. Note that this decay length is a
function of the electric field \emph{and} the doping level. The decay length of
$\lambda=2.4\,\text\AA$ corresponds well to experimentally found values for similar gate
levels \cite{Ohta2007}. We have also tested this on fewer layers with consistent results.

% \subsubsection{Buffer atoms}
% \label{ssec:buffer}
% \input{buffer.tex}

% \subsubsection{The density at the electrode boundary}
% \label{ssec:bulk-DM}
% \input{electrode-boundary.tex}

\subsection{Charge conservation}
\label{ssec:dEf}

A recurring issue with NEGF calculations is excess charge compared to a charge neutral
device region. Especially, for weakly screening systems and for non-equilibrium the SCF
loop may converge slowly and with great difficulty due to larger deviations from charge
neutrality in the beginning of the SCF loop \cite{Engelund2016}. To remedy this we have
implemented an option which introduces a shift in the potential inside the device region,
$\cd\E$, which sometimes can push the SCF loop towards the self-consistent solution
with a charge neutral device region or that the assumption of equilibrium in the electrodes
%
%
% We denote any excess charge in a converged calculation as $\delta q$. It arises due to the
% change in boundary conditions, from periodic to an open boundary condition where a bulk
% electrode, and thus bulk density of states. Hence if the simulation cell was extended in
% the semi-infinite directions one should eventually retrieve a $\delta q\sim0$ as the
% extended electrodes screen the charges. Another important parameter is the Hartree
% potential which is solved using FFT, thus inherently adding a charged background for any
% excess charge. The excess background is equivalent to $-\delta q$. Needless to say a too
% high $\delta q$ can in certain situations create an un-physical potential profile.
%
% In general for EGF calculations should the scattering region screen out the charge
% redistribution such that no excess charge exists. If any excess charge exists it is an
% indication of poor integration parameters and/or too small scattering region which
% prohibits proper screening. 
%
% In cases where the charge neutrality is not enforced care has to be taken. \tsiesta\
% enables an empirical correction which forces $\delta q\to0$. 

The potential shift is obtained from the requirement that the net charge of the device region,
$q_D$, should be zero. We use the total device DOS, $D(\E_R)$, at an initial energy
reference level, $\E_R$, located in the bias window,
\begin{equation}
  \label{eq:correction:Fermi}
  q_{D} = \delta q+D(\E_R)\,\cd\E = 0\quad \Rightarrow\quad  \cd\E= -\frac{\delta q}{D(\E_R)}\,.
\end{equation}
We assume here that the DOS vary slowly on the scale of the bias and $\kT$ such that the
reference energy is not critical. To lowest perturbation order in the potential shift we
get a change in the density matrix in terms of the spectral density matrix,
$\Delta\DM = \cd\E\,(d\DM(\E_R)/d\epsilon)$, which when added to $\DM$ in the SCF loop
enforces a charge neutral device region. During the SCF-loop we then update the reference
energy $\E_R\leftarrow\E_R+\cd \E$. When the SCF-loop has converged we can check the
degree of charge neutrality $\delta q$ and potential shift, $\cd\E$. These should both be
small such that the feature can be turned off in a restarted calculation and converge
without being invoked. If this is not possible it may indicate that the device region has
a too short screening region towards the electrodes or, for high bias, that the
approximation of equilibrium density and potential in the electrodes is not adequate.

% \subsubsection{NetCDF-4 support}
% \label{ssec:ncdf}
% \input{netcdf4.tex}

\subsection{Thermoelectric effects under NEGF}
\label{ssec:thermo}

To the authors knowledge, thermoelectric effects are currently only studied under
equi-temperature distributions for NEGF calculations. However, in principle such effects
require population statistics to be correctly described using self-consistent NEGF
calculations. Our implementation naturally permits such generality as the
chemical potential and electronic temperature can be set independently for each electrode.

\begin{figure}
  \centering
  \includegraphics{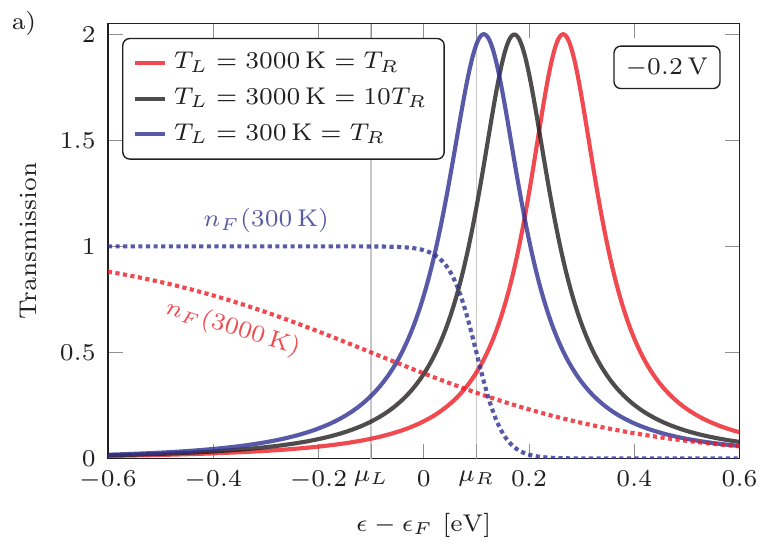}\quad
  \includegraphics{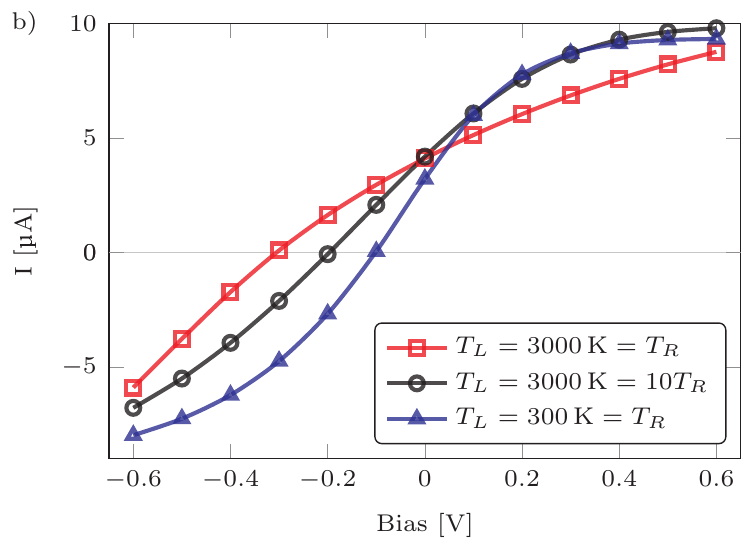}
  \caption{Example of a thermoelectric calculation for a simple 1D setup consisting of a central C atom weakly coupled to two perfect semiinfinite C-wires. %
      a) Impact on the electronic
      transmission function $\TT(\E,V=\unitr{-0.2}{eV})$ of the electrode temperatures $T_{L,R}$.
      The Fermi functions are also included for clarity of the difference in population. 
      b) $IV$ characteristics computed from three distinct NEGF calculations (uniform
      temperature $T=300$, $\unitr{3000}K$ vs temperature difference $T_L=\unitr{3000}K$ and
      $T_R=\unitr{300}K$)
      as detailed in the text.
      \label{fig:thermoelectric}
  }
\end{figure}

As an example of how different electronic temperatures in the electrodes can impact the
electronic structure in the device region we have performed calculations for a simple 1D
setup consisting of a central C atom weakly coupled to two semi-infinite, 1D
C-wires (lattice constant $a=\unit{1.30}{\text{\AA}}$). The C-C distance between the central atom
and the electrodes was set to $d= \unit{2.50}{\text{\AA}}$. According to the band structure of the
electrodes (not shown), the $2p_x$ and $2p_y$ orbitals on the equidistant lattice sites form a degenerate, half-filled
band, which couples to the $2p_x$ and $2p_y$ orbitals of the central C-atom. This
situation leads to the degenerate resonance structure in the transmission function as
shown in \fref{fig:thermoelectric}a. Note that the position of this transmission
resonance, i.e., the energetic position of the C-atom $2p_x$ and $2p_y$ orbitals, varies
substantially for the three considered choices of the electrode temperatures.  The
thermoelectric calculations for the electron current shown in \fref{fig:thermoelectric}b
correspond to the situation in which the electronic temperature of the left (right)
electrode is fixed at $T_L=\unitr{3000}K$ ($T_R=\unitr{300}K$) in the Landauer formula,
but where different temperature settings are used in the underlying self-consistent
DFT-NEGF calculation as detailed in the figure legend.

These results exemplify that in situations with temperature differences between the
electrodes, the fully self-consistent $IV$ characteristics (black curve in
\fref{fig:thermoelectric}b) cannot be determined using a uniform temperature (red or
blue curves in \fref{fig:thermoelectric}b).
While the extreme temperature difference and narrow resonance at play in this example may
seem far away from practical situations, we emphasize that it is generally desirable to be able to include this temperature effect at no additional computational cost. We further speculate that these methods could find
relevance to describe situations with effective electronic temperatures substantially above the lattice temperature, e.g., as originating from optical driving \cite{JoUlCi.13.DirectViewHot,GiPeMi.13.Snapshotsnonequilibrium}.

\section{Green function transport and techniques}
\label{sec:tbtrans}

As a post-processing tool for DFT-NEGF calculations of the self-consistent Hamiltonian, we
have also developed the next-generation \tbtrans\ (and its offshoot \phtrans). \tbtrans\
enables the calculation of electronic transport, electronic thermal energy transport,
phonon transport (\phtrans), and provide analysis tools such as the (spectral) density of
states, the sub-partition eigenstate projected transmission (molecular or phonon
eigenmode), interpolated $I$-$V$ curves, transmission eigenvalues and
orbital/bond-currents \cite{Todorov2002}. Thus it is possible obtain thermoelectric
quantities such as the Seebeck and Peltier coefficients as well as calculation of
thermoelectric figure-of-merit including lattice heat-transport in the harmonic
approximation. All features are enabled for general multi-electrode ($N_\idxE\ge 1$)
setups. The features discussed below -- transmission functions, (spectral) density of
states, sub-partition eigenstate projected transmissions, transmission eigenvalues, and
bond-currents -- encompass both the electronic and phononic versions, but for brevity we
refer only to the electronic part in the following.

Besides input from DFT calculations \tbtrans\ is also able to handle general
user-created tight-binding models or by using Hamiltonians from other sources (Wannier
basis, etc.). Thus \tbtrans\ is now a stand-alone application capable of being used for
large-scale, ballistic transport calculations. For example, simulations of square graphene
flakes exceeding $10^6$ atoms may easily be done on typical office
computers (one orbital/atom). An interface on top of \tbtrans\ for creating corrections to DFT -- such as
scissor, LDA+U, magnetic fields, etc. -- has also been created such that the capabilities of
\tbtrans\ are \emph{not} restricted by the developers, but, in principle, by the user. This
interface also allows the extraction of Hamiltonians from other programs to the \tbtrans\
format. Other efficient methods involve similar constructs \cite{Mason2011,Ferrer2014}.

\subsection{Transmission function for \texorpdfstring{$N_\idxE$}{Ne} electrodes}
The transmission functions can be calculated using the scattering matrix formalism and
obtained from the Green function using the generalized Fisher-Lee relation
\cite{Fisher1981,sanvito,Yeyati2000} (or the Lippmann-Schwinger equation
\cite{Lippmann1950,Cook2011}). The elements of the scattering matrix, at a given $\kk$, can be written as
\begin{equation}
  \label{eq:tbt:scattering}
  s_{\idxE\idxE',\kk} = -\delta_{\idxE\idxE'}\mathbf I +i \Scat_{\idxE,\kk}^{1/2}\G_\kk^{\phantom{/}}\Scat_{\idxE',\kk}^{1/2},
\end{equation}
where $\idxE$ and $\idxE'$ refer to two electrodes and $\delta_{\idxE\idxE'}$ is the
Kronecker delta. We implicitly assume energy dependence on all quantities. The
transmission (probability) from electrode $\idxE$ to $\idxE'$ is
\begin{equation}
  \label{eq:tbt:T}
  \TT_{\idxE\idxE',\kk}=\Tr\big[s^\dagger_{\idxE\idxE',\kk} s^{\phantom{\dagger}}_{\idxE\idxE',\kk}\big]
  =\left\{
    \begin{aligned}
      &\Tr\big[\G_\kk^{\phantom{\dagger}}\Scat^{\phantom{\dagger}}_{\idxE,\kk}\G^\dagger_\kk\Scat^{\phantom{\dagger}}_{\idxE',\kk}\big], 
      & \text{for $\idxE\neq\idxE'$}
      \\
      &\RE_{\idxE,\kk}, 
      & 
      \text{for $\idxE=\idxE'$}
    \end{aligned}
  \right. 
\end{equation}
where reflection (probability) is defined as
$\RE_{\idxE,\kk} \equiv \TT_{\idxE\idxE,\kk}$.  
It is instructive to write the aggregate transmission $\TT_{\idxE,\kk}$ out of an
electrode $\idxE$ (see Ref.~\cite{Lake1997}) and the reflection $\RE_{\idxE,\kk}$ as
\begin{align}
  %\label{eq:tbt:T}
  %\TT_{\idxE\idxE'} &=
  %\Tr\big[\G\Scat_{\idxE}\G^\dagger\Scat_{\idxE'}\big],\quad\text{for $\idxE\neq\idxE'$}
  %\\
  \label{eq:tbt:T-out}
  \TT_{\idxE,\kk} &\equiv\sum_{\idxE'\neq\idxE}\TT_{\idxE\idxE',\kk} 
  =
  i \Tr\big[(\G_\kk^{\phantom{\dagger}}-\G_\kk^\dagger)\Scat_{\idxE,\kk}^{\phantom{\dagger}}\big]
  -\Tr[\G_\kk^{\phantom{\dagger}}\Scat_{\idxE,\kk}^{\phantom{\dagger}}\G_\kk^\dagger\Scat_{\idxE,\kk}^{\phantom{\dagger}}],
  \\
  \label{eq:tbt:R}
  \RE_{\idxE,\kk} &= M_{\idxE,\kk} -\TT_{\idxE,\kk} .
  %=  M_\idxE-\Big\{
%   i \Tr\big[(\G-\G^\dagger)\Scat_\idxE\big]
%  -\Tr[\G\Scat_\idxE\G^\dagger\Scat_\idxE]
%  \Big\}.
\end{align}
The reflection is here conveniently written as a difference between the bulk electrode
transmission $M_\idxE$ (i.e., number of open channels/modes in electrode $\idxE$ at the
given energy) and the aggregate transmission $\TT_\idxE$ (scattered part into the other
electrodes).
%\note{Since two waves are involved in the reflection (incoming and
%    scattered wave) the resulting probability current contains an interference term
%    between these (second term in Eq.~\ref{eq:tbt:R}). (I do not quite get this, isn't the
%    above two lines enough?)}
%
From Eqs.~\eqref{eq:tbt:T}, \eqref{eq:tbt:T-out} and \eqref{eq:tbt:R} one may easily prove
the transmission equivalence $\TT_{\idxE\idxE',\kk}\equiv\TT_{\idxE'\idxE,-\kk}$ as well as
$\TT_{\idxE\idxE',\kk}=\TT_{\idxE\idxE',-\kk}$ based on time-reversal symmetry.
% * <nicolas_lorente001@ehu.eus> 2016-07-13T16:20:46.122Z:
%
% shouldn't idxE and idxE' be exchanged when kk goes to -kk??
% Yes, but one may actually easily prove that
%  \TT_{\idxE\idxE',\kk}=\TT_{\idxE\idxE',-\kk}
% when time-reversal symmetry applies. And this result is easily
% obtained using the above equations
%
% ^.
% * <nicolas_lorente001@ehu.eus> 2016-07-13T16:20:31.715Z:
%
% ^.
%
Equation~\eqref{eq:tbt:T-out} displays an important, and often overlooked detail. In transport
calculations with $N_\idxE=2$, one can calculate the transmission using only a
sub-diagonal part of the Green function and only one scattering matrix. We stress that the
quantities calculated may have numerical deficiencies as $\Tr[(\G-\G^\dagger)\Scat_\idxE]$
and $\Tr[\G\Scat_\idxE\G^\dagger\Scat_\idxE]$ may both be numerically large which leads to
inaccuracies when the transmission is orders of magnitudes smaller than the
reflection\footnote{Typically this is a problem for systems with relatively large bulk
    transmissions compared to the transmission. If the number of incoming channels are
    only a couple of magnitude orders larger than the transmission the numerical precision
    is adequate.}.  

Further, the transmission may be split into transmission eigenvalues, 
\begin{equation}
  \label{eq:tbt:Teig}
  \TT_{\idxE\idxE',\kk} = \sum_i\TT_{i,\idxE\idxE',\kk}\,,
\end{equation}
where $\TT_{i,\idxE\idxE',\kk}$ are the eigenvalues of the column matrix
$\G_\kk\Scat_{\idxE,\kk}\G_\kk^\dagger\Scat_{\idxE',\kk}$ ($\idxE\neq\idxE'$)
\cite{Paulsson2007a}. Due to the matrix product being a column matrix ($\Scat_\idxE$ have
a limited extend due to the LCAO basis) the eigenvalue calculation can be reduced
substantially by realizing the following equation,
\begin{equation}
  \det\big(\G_\kk^{\phantom{\dagger}}\Scat_{\idxE,\kk}^{\phantom{\dagger}}\G_\kk^\dagger\Scat_{\idxE',\kk}^{\phantom{\dagger}} -\lambda\ID\big)= 
  \det\Big(
  \begingroup
  \def\arraystretch{0.}%  1 is
  \setlength\tabcolsep{0pt}
  \begin{tabular}{|@{}ccc@{}|}
    \hline
    \color{white}\sq& & \sq
    \\
    & \color{white}\sq & \sq
    \\
    &  & \sq
    \\
    \hline
  \end{tabular}
  \endgroup  -\lambda\ID
  \Big)
  = 
  \det\left(
  \begin{bmatrix}
    \mathbf 0 & \mathbf 0 & \mathbf A
    \\
    \mathbf 0 & \mathbf 0 & \mathbf B
    \\
    \mathbf 0 & \mathbf 0 & \mathbf C
  \end{bmatrix} -\lambda\ID\right)\to\det(\mathbf C - \lambda\ID).
\end{equation}
The transmission eigenvalues are for instance important when calculating Fano factors describing shot noise \cite{Schneider2012}.

Once the transmission function is calculated we can calculate the electrical current $I_{\idxE\idxE'}$ and thermal
energy transfer $Q_{\idxE\idxE'}$ as 
% I think idxE -> idxE' should be implicit here (already added above)
\begin{align}
  \label{eq:J}
  I_{\idxE\idxE'} &= \frac{\mathrm{G}_0}{2|e|}\iint_\BZ\dEBZ\cd \kk\dd\E
  \, \TT_{\idxE\idxE',\kk}(\E) \big[n_{F,\idxE'}(\E) - n_{F,\idxE}(\E)\big],
  \\
  \label{eq:H}
  Q_{\idxE\idxE'} &= \frac1h\iint_\BZ\dEBZ\cd \kk\dd \E \, 
  \TT_{\idxE\idxE',\kk}(\E)(\E-\mu_{\idxE}) \big[n_{F,\idxE'}(\E) - n_{F,\idxE}(\E)\big],
\end{align}
where $\mathrm{G}_0=2e^2/h$ is the conductance quantum (spin-degenerate case). When
time-reversal symmetry applies one has the relations $I_{\idxE\idxE'}= -I_{\idxE'\idxE}$
and
$Q_{\idxE\idxE'}+Q_{\idxE'\idxE}= (\mu_\idxE'-\mu_{\idxE})/|e|
I_{\idxE\idxE'}\equiv W$.
The latter expresses that the net work done, $W$, equals the net heat supplied.

% \revision{TF comment: If $\mu_\idxE>\mu_{\idxE'}$ (and $T_\idxE=T_{\idxE'}= 0$) we have $I_{\idxE\idxE'}<0$ (and $I_{\idxE'\idxE}>0$) (net electrons move from $\idxE$ to $\idxE'$ but the electrical current flows in the opposite direction). We also have $Q_{\idxE\idxE'}>0$ (and $Q_{\idxE'\idxE}<0$). The work is $W=(\mu_{\idxE'}-\mu_{\idxE})/\revision{|e|} I_{\idxE\idxE'}>0$. In other words, $\Delta Q = Q_{\idxE\idxE'}+Q_{\idxE'\idxE}=W>0$ --- OK, I think the signs make sense!}
% Proof (TF)
% \begin{align}
% Q_{\idxE\idxE'} &= \frac1h\iint_\BZ\dEBZ\cd \kk\dd \E \, 
%   \TT_{\idxE\idxE',\kk}(\E)(\E-\mu_{\idxE}) \big[n_{F,\idxE'}(\E) - n_{F,\idxE}(\E)\big]\nonumber\\
%   &= \frac1h\iint_\BZ\dEBZ\cd \kk\dd \E \, 
%   \TT_{\idxE\idxE',\kk}(\E)(\E-\mu_{\idxE'}+\mu_{\idxE'}-\mu_{\idxE}) \big[n_{F,\idxE'}(\E) - n_{F,\idxE}(\E)\big]\nonumber\\
%   &= \frac1h\iint_\BZ\dEBZ\cd \kk\dd \E \, 
%   \TT_{\idxE\idxE',\kk}(\E)(\E-\mu_{\idxE'})\big[n_{F,\idxE'}(\E) - n_{F,\idxE}(\E)\big]
%   +(\mu_{\idxE'}-\mu_{\idxE})\frac1h\iint_\BZ\dEBZ\cd \kk\dd \E \, 
%   \TT_{\idxE\idxE',\kk}(\E)\big[n_{F,\idxE'}(\E) - n_{F,\idxE}(\E)\big]\nonumber\\
%   &= -Q_{\idxE'\idxE}+ \frac{\mu_{\idxE'}-\mu_{\idxE}}{e} I_{\idxE\idxE'}
% \end{align}

We note that one may use efficient interpolation schemes for the BZ averages to reduce the
required number of $\kk$-points, which is typically much larger than the $\kk$-sampling
necessary for the corresponding density matrix calculation \cite{Falkenberg2015}.

\subsection{Inversion algorithm --- again}
\begin{figure}
  \centering
  \includegraphics{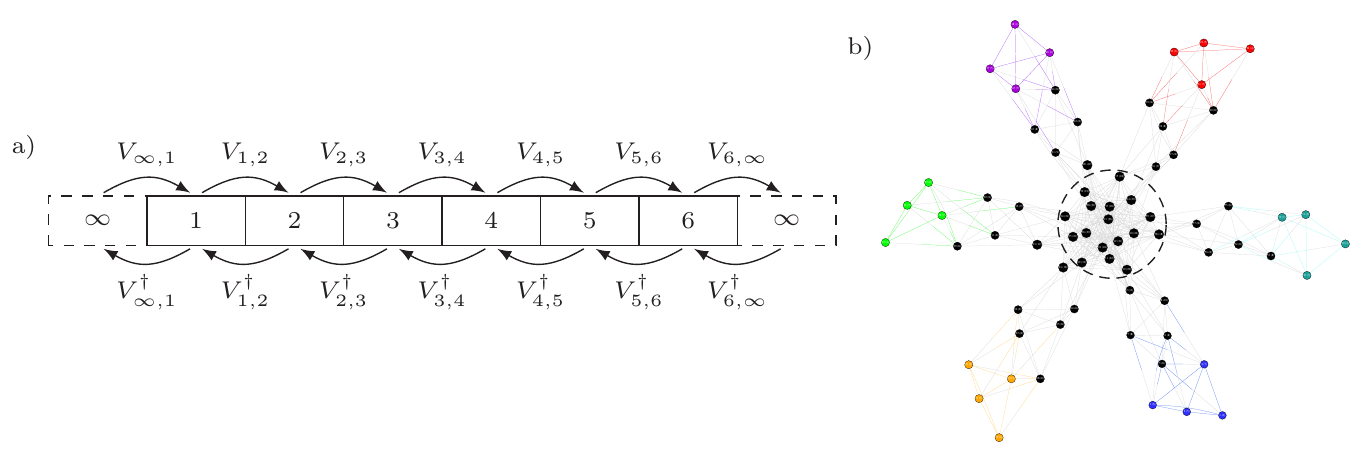}
  \caption{The BTD algorithm in \tbtrans. a) Partitioning of a standard two-terminal setup
      for the recursive Green function method (BTD). b) Connectivity graph for the
      $N_\idxE=6$ terminal device used in \protect\fref{fig:charge} where each dot
      represents an atom (non-black colored atoms correspond to an electrode) and every
      line represents one or multiple connections between the two atoms. Instead of
      calculating the Green function in the whole device space one can shrink the problem
      to a smaller region as long as the electrode branches do not couple directly to each
      other. An example region is shown with a dashed circle.
      \label{fig:tbt:Gf}}
\end{figure}

The Green function algorithm in \tbtrans\ is similar to the one in \tsiesta\ (but not the
same). In Figs.~\ref{fig:system-setup} and \ref{fig:tbt:Gf} we exemplify the method. In
\fref{fig:tbt:Gf}a we show a regular two-terminal device with $8$ partitions. In the BTD
method one can down-fold the self-energy from the left $\infty$ block up till e.g. the
block $6$, using \eref{eq:BTD:algo}, or as explained in
Refs.~\cite{Godfrin1991,Petersen2008}. Then the current is calculated using the standard
\eref{eq:J} based on the transmission evaluated from the down-folded quantities
calculating the Green function \eref{eq:BTDs} and self-energies in the sub-space of
block $6$. However, one could equally have chosen block $3$, or the combined blocks $3$
and $4$.  The advantage of this abstraction is threefold: 1) the Green function is
obtained in the chosen blocks of interest only, 2) choosing fewer blocks reduces the
computational complexity which greatly speeds up the calculation, and 3) choosing small
blocks reduces the required memory which enables extreme scale calculations. From 1) it
follows that quantities such as local density of states can be calculated at arbitrary
positions in the calculation cell by selecting specific blocks. However, we note that for
non-orthogonal basis sets an increased block including the overlap region is required in
order to obtain LDOS, Mulliken charges, etc. On the other hand, if one is only
interested in the transmission/current one can resort to choosing the smallest block to
achieve the highest throughput \cite{Petersen2008}.

Similarly, down-folding of the self-energies can also be achieved in an $N_\idxE$-terminal
device. As an example, a 6-terminal system (e.g., the setup in \fref{fig:tbt:Gf}b)
can be split into several blocks with their own downfolding of self-energies.
One chooses a region $D$ (see \fref{fig:system-setup}) and the extended electrode regions
may be used to down-fold the self-energy. However, one may not choose $D$ such that either
of the electrodes are directly coupled, since this choice would entangle the self-energies
and their origin would be lost.

After selecting the region $D$, \tbtrans\ creates $N_\idxE+1$ different BTD matrices for optimal
performance. These are $N_\idxE$ BTD matrices for each electrode including the down-folding region ($\{\idxE_i,\idxE_i+\}$
in \fref{fig:system-setup}), and one additional BTD matrix for region $D$. Each BTD matrix uses
its own pivoting scheme to reduce the bandwidth and increase performance.  Due to the
pivoting, the Hamiltonian structure cannot be generically outlined in matrix format. Yet
it can be formulated equivalently as in Ref.~\cite{Ferrer2014} with the possibility of
letting the user define the ``extended scattering region''. \tbtrans\ allows the user to do this via atomic indices, and hence no knowledge of the algorithms
underlying the pivoting and other details are needed. 
Lastly, we note that \tbtrans\ is also implemented using OpenMP 3.1 threading and scales
like shown in \fref{fig:perf}c for large systems.

\subsection{\texorpdfstring{$I-V$}{I-V} curves using Hamiltonian interpolation}
Although the present work represents a significant leap in performance of \tsiesta\ and
\tbtrans, self-consistent NEGF calculations are still heavy, especially when
current-voltage ($I-V$) characteristics with many bias points are required. To ease, and
quite accurately, calculate $I-V$ curves we have implemented an interpolation scheme based
on $N_V$ separate NEGF calculations at different bias conditions.
% As far as I can see from the ATK manual they perform interpolation of I-V points, so
% nothing anybody can't do themselves... :)
If $N_V=2$ we do a linear interpolation/extrapolation of the Hamiltonian, and if $N_V>2$ a
spline interpolation is also possible. Note that a spline extrapolation is equivalent,
unsurprisingly, to linear extrapolation.
\begin{figure}
  \centering
  \includegraphics{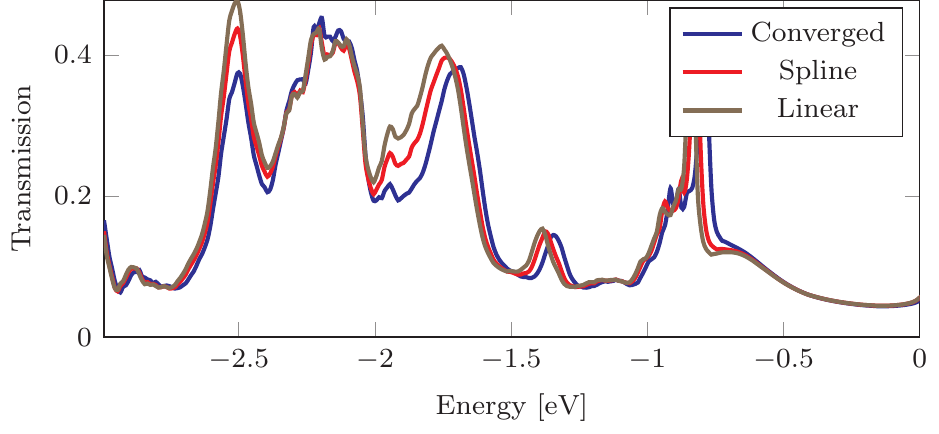}
  \caption{Interpolation of the transmission function for a Cu-tip--\Csix/Cu(111) junction using either spline or linear
      interpolation of the Hamiltonian at $V=\unitr{-1.5}V$ using the converged Hamiltonians
      for $\unitr{-2}V$, $\unitr{-1}V$, $\unitr{0}V$, $\unitr{1}V$ and $\unitr{2}V$. The spline
      interpolation (red curve) agrees significantly better with the self-consistent solution (blue curve) than the linear interpolation (brown curve).
      \label{fig:tbt:interp}
  }
\end{figure}
As a test example we consider a molecular contact system consisting of a periodic array of
\Csix\ molecules buried in the first surface layer of a Cu(111) surface contacted by a
tip. This example is taken from Ref.~\cite{Schneider2015}. In \fref{fig:tbt:interp} we
compare the $\kk$-averaged transmission functions based on the self-consistent (converged)
Hamiltonian at a bias of $\unitr{-1.5}V$ and the corresponding one interpolated from
self-consistent Hamiltonians at $\unitr{-2}V$, $\unitr{-1}V$, $\unitr{0}V$, $\unitr{1}V$
and $\unitr{2}V$ (linear interpolation from the two closest bias points and spline using
all points). In this example the bias point $\unitr{-1.5}{V}$ was ``worst case scenario''
where the exact and interpolated results deviated the most out of 47 interpolations and
extrapolations in the range $\unitr{-2.4}V$ to $\unitr{2.4}V$ (in equal steps of
$\unitr{0.1}V$). Although both interpolation schemes perform very well, the spline
interpolation clearly outperforms the linear interpolation, retaining a better agreement
with the self-consistent calculation. This also works for $N_\idxE\neq2$ if the chemical
potentials are linearly dependent.

\subsection{\tbtrans\ as transport back-end and feature generalization}
Even though \tbtrans\ is developed with \tsiesta\ in mind, its use is far from restricted
to this. \tbtrans\ implements flexible NetCDF-4 support, and the Hamiltonian can thus
alternatively be supplied through a NetCDF-4 file. This makes \tbtrans\ accessible as a
generic transport code without a need for lower-level Fortran coding and/or knowledge of
the \siesta\ binary file format. To accommodate such so-called ``tight-binding''
calculations, we have developed a LGPL licensed Python package \sisl\ \cite{sisl} which
facilitates an easy interface to create large-scale (non-)orthogonal tight-binding models
for arbitrary geometries. This package is a generic package with further analysis tools
for \siesta\ such as file operations and grid operations (interaction with fdf-files,
\vasp\ files, and density grids, potential grids, etc.).
\sisl\ allows any number of atoms, different orbitals per atom, (non-)orthogonal basis
sets, mixed species in a 3D super-cell approach. Furthermore, as \sisl\ is based on Python
it allows abstraction such that unnecessary details of the complex data structures are
hidden from the users. Currently it is interfaced to read parameters from \siesta, \gulp\
and \textsc{Wannier90} \cite{Gale2003,Mostofi2014}.

\begin{figure}
  \centering
  \begin{minipage}[c]{0.45\linewidth}
    \footnotesize
    \begin{verbatim}
import sisl
bond = 1.42
graphene = sisl.geom.graphene(bond)
# Create a 100x100x2 = 20000 atom graphene flake
flake = graphene.repeat(100, axis=0).tile(100, 1)
H = sisl.Hamiltonian(flake)
#       U          t1
dR = [0.1 * bond, 1.1 * bond]
for ia in flake:
    idx_a = flake.close(ia, dR=dR)
    # Define Hamiltonian elements
    H[ia,idx_a[0]] =  0.   # on-site
    H[ia,idx_a[1]] = -2.7  # nearest neighbor
H.write('FLAKE.nc')
    \end{verbatim}
  \end{minipage}
  \quad
  \begin{minipage}[c]{0.45\linewidth}
    \footnotesize
    \begin{verbatim}
import sisl
bond = 1.42
graphene = sisl.geom.graphene(bond)
flake = graphene.repeat(100, axis=0).tile(100, 1)
# Remove atoms in a circular region to create a hole
hole = flake.remove(flake.close(flake.center(), dR=10*bond))
H = sisl.Hamiltonian(hole)
dR = [0.1 * bond, 1.1 * bond]
for ia in hole:
    idx_a = hole.close(ia, dR=dR)
    # Define Hamiltonian elements
    H[ia,idx_a[0]] =  0.   # on-site
    H[ia,idx_a[1]] = -2.7  # nearest neighbor
H.write('HOLE.nc')
    \end{verbatim}
  \end{minipage}

  \caption{\sisl\ code. Left: Creation of a periodic graphene flake with 20,000 atoms with
      nearest neighbor interactions ($\E_0=0$ and $\E_1=-2.7$). 
      Right: Creation of a periodic graphene flake with a hole having a
      diameter of $20$ bond lengths.
      \label{alg:tbt:python}
  }
  
\end{figure}
In \fref{alg:tbt:python} we list two \sisl-code examples. The left example creates a simple $20,000$ atom graphene flake with nearest neighbor interaction. The right example creates the same graphene structure but with a hole with a diameter of $20$ times the bond
length at the center of the flake. 
In the SM we have added several example scripts for graphene with
various tight-binding parameter sets \cite{Hancock2010} as well as an example of how to
create a tight-binding transport calculation.

\subsubsection{Feature generalization -- \texorpdfstring{$\delta\HH$}{dH}}
\label{ssec:tbt:dH}
Green function codes are often limited by the implemented features. Yet a large scope of
features can be described using a \emph{correction} of the Hamiltonian elements due to
local smearing, scissor operators, magnetic fields, the SAINT method
\cite{Garcia-Suarez2011}, etc. All in all they can be summarized in this one equation
\begin{equation}
  \label{eq:tbt:dH}
  \HH_\kk \leftarrow \HH_\kk + \delta\HH_\kk(\E),
\end{equation}
where $\delta\HH_\kk(\E)$ encompass any correction for the Hamiltonian, whatever that
may be. $\delta\HH_\kk(\E)$ may be of a real quantity, or a complex quantity, depending on
its physical origin.

Instead of letting the developers decide which features should be implemented we enable
users to create their own features by altering the Hamiltonian elements, however they wish,
simply by creating the $\delta\HH_\kk(\E)$ correction. \sisl\ \cite{sisl} efficiently creates the $\delta\HH_\kk(\E)$ matrix without any prior knowledge of Fortran
or the data format for \tbtrans. We note that \sisl\ is capable of reading the DFT-NEGF
self-consistent Hamiltonian and edit it directly in Python.

The $\delta\HH_\kk(\E)$ comes in four variants to control different parts of the
calculation. 
\begin{enumerate}
  \item $\delta\HH$ with no energy nor $\kk$-point dependencies,
  \item $\delta\HH_\kk$ with only $\kk$-point dependency,
  \item $\delta\HH(\E)$ with only energy dependency,
  \item $\delta\HH_\kk(\E)$ with both $\kk$-point and energy dependencies.
\end{enumerate}
This \emph{feature} enables a broader population of users to contribute with functionality
to a public code, and we encourage contributions to the \siesta\ mailing list as well as
the \sisl\ GitHub repository which may be used as a base for further development of
features for the transport code \tbtrans.

\subsection{Molecular state projection transmission}
\label{ssec:tbt:MPSH}
\newcommand\set[1]{\{\!#1\}}
\newcommand\kbpkb[4]{\ket{#1}\bpk{#1}{#2}{#3}{#4}\bra{#2}}
\newcommand\bpk[4]{\overset{#3,#4}{\bk{#1}{#2}}}

There are several approaches to analyze transport properties besides considering the local
density of states. One may decompose the transmission into eigenchannels and plot the
corresponding scattering states \cite{Paulsson2007a} or the
bond-currents \cite{Solomon2010}. However, this does not necessarily give a clear answer to
the basic question as to which of the eigenstates in the central region takes part in
transport, e.g., which molecular levels transmit the electrons in a molecular contact
between metals.  \tbtrans\ allow a very flexible analysis of the transport through such
eigenstates.  In the following we use molecular eigenstates as the terminology and have a
central molecule as ``bottle-neck'' in mind, however, the method is not limited to
this type of setup.

We can define a sub-space of the full device region consisting of the same or fewer basis
components $\set M$ (henceforth referred to as ``molecule'').  The Hamiltonian for this
region have previously been denoted Molecular Projected Self-consistent
Hamiltonian (MPSH) \cite{Stokbro2003c} with corresponding eigenstates,
\begin{equation}
  \HH_{\set M}\ket{M'_i} = \eig_i^{\set M}\SO_{\set M} \ket{M'_i},
\end{equation}
where $\ket{M'_i}$ are the generalized eigenvectors defined in the basis functions of the
non-orthogonal basis. In order to create orthogonal eigenvectors ($\ket{M_i}$) we rotate
the basis set to form orthonormal projection vectors using the L\"owdin
transformation \cite{Lowdin1950}.

% We note that the vectors in the eigenset in $\set M$ will only coincide with
% eigenvectors in the eigenset $\set S$ if the subset $\set M$ is fully decoupled from the
% complement basis functions $\set C=\set S\setminus\set M$. In cases where the molecule
% is only slightly hybridized by $\set C$, one can assume that the molecular eigenstates
% are sufficiently describing a subset of the system eigenstates. Only close to this limit
% can the transmission through eigenstate orbitals be adequately attributed.

Inserting the complete $\set M$ (orthogonalized) basis in the expression for the transmission
(assuming $\kk$ and $\E$ dependencies implicit, and
$\{\idxE,\idxE'\}=\{L,R\}$) we get
\begin{align}
  \TT_{LR}&=\Tr[\G\Scat_L\G^\dagger\Scat_R] 
  \\
  &=\Tr\Big[
   \G
     \sum_j\ket{M_j}\bra{M_j}\Scat_{L}\sum_{j'}\ket{M_{j'}}\bra{M_{j'}}
   \G^\dagger
     \sum_i\ket{M_i}\bra{M_i}\Scat_{R}\sum_{i'}\ket{M_{i'}}\bra{M_{i'}}
   \Big].
\end{align}
The matrix element of the broadening matrix, $\bra{M_j}\Scat_{L}\ket{M_j'}$, describes the coupling of the MPSH states to
electrode $L$ and how these are mixed/hybridized due to this.  How such a projector is
chosen is outside the scope of this article, but we refer to Ref.~\cite{Rangel2015} for
additional projectors. In \tbtrans\ we save all scalar quantities
$\bra{M_j}\Scat_{L}\ket{M_{j'}}$ for an extra level of information. It also allows for
ways to break the total transmission into components for each MPSH state in a very
flexible way as illustrated in the following.

\begin{figure}
  \centering
  \includegraphics{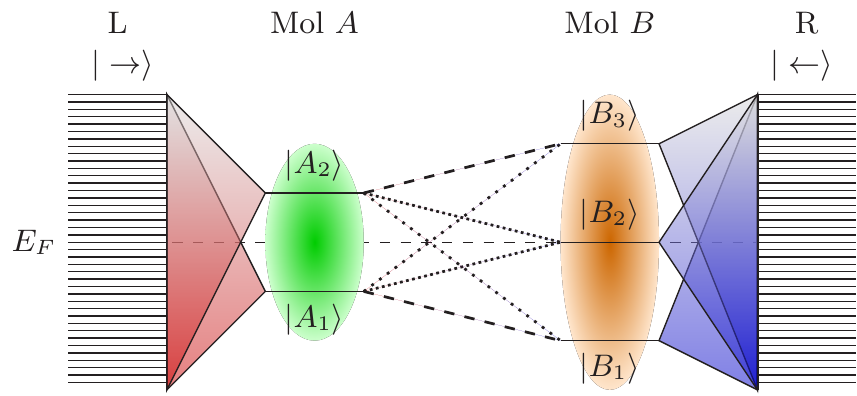}
  \caption{Schematic illustration of two molecules coupled to 2 electrodes. One can follow each of
      the lines connecting molecule $A$ and $B$. To scatter into the right electrode they have to
      scatter across the molecular states $\ket{A_1}$, $\ket{A_2}$, $\ket{B_1}$, $\ket{B_2}$ and $\ket{B_3}$.
      \label{fig:tbt:MPSH}
  }
\end{figure}

As an example we consider schematic in \fref{fig:tbt:MPSH} corresponding to electron transport through a ``bridge'' consisting of two molecules ($A$ and $B$) with 2 and 3 MPSH characteristic eigenstates each, respectively.  It is then possible to extract the transmission probability corresponding to, say, electrons injected from $L$ into state $\ket{A_1}$ and extracted to $R$ via the set
$\ket{B_{\set{1,2}}}$, yielding
\begin{equation}
  \label{eq:tb:projection:A-B}
  \TT_{A_{\set{1}}B_{\set{1,2}}}
  =\Tr\Big[
  \G
  \ket{A_1}\bra{A_1}\Scat_{L}\ket{A_1}\bra{A_1}
  \G^\dagger
  \sum_{j=1}^2\ket{B_j}\bra{B_j}\Scat_{R}\sum_{j'=1}^2\ket{B_{j'}}\bra{B_{j'}}
  \Big].
\end{equation}
We note that such projected transmissions may be larger than the non-projected total
transmission due to interference effects.
Additionally this projection scheme also allows investigation of the difference between projections on
incoming-outgoing scattering states. That is, for a single molecular junction ($A$ with 2
molecular states) one may
calculate ($\ID=A_{\set{1,2}}$):
\begin{subequations}
  \label{eq:tb:proj}
\begin{align}
  \label{eq:tb:proj-I}
  \TT_{A_{\set{1}}\ID}
  &=\Tr\Big[
  \G
  \ket{A_1}\bra{A_1}\Scat_{L}\ket{A_1}\bra{A_1}
  \G^\dagger
  \Scat_{R}
  \Big], &[\rightarrow]
  \\
  \label{eq:tb:proj-O}
  \TT_{\ID A_{\set{1}}}
  &=\Tr\Big[
  \G
  \Scat_{L}
  \G^\dagger
  \ket{A_1}\bra{A_1}\Scat_{R}\ket{A_1}\bra{A_1}
  \Big],&[\leftarrow]
  \\
  \label{eq:tb:proj-IO}
  \TT_{A_{\set{1}}A_{\set{1}}}
  &=\Tr\Big[
  \G
  \ket{A_1}\bra{A_1}\Scat_{L}\ket{A_1}\bra{A_1}
  \G^\dagger
  \ket{A_1}\bra{A_1}\Scat_{R}\ket{A_1}\bra{A_1}
  \Big],&[\leftrightarrow]
\end{align}
\end{subequations}
where $\rightarrow$ and $\leftarrow$ refers to incoming and outgoing projections,
respectively, and $\leftrightarrow$ refers to a simultaneous projection of incoming and
outgoing. These 3 transmissions are generally not equal due to asymmetric coupling and/or
hybridization with the electrodes. Lastly, we allow the projection states to be both
$\kk$-resolved or $\Gamma$-point only. If the MPSH eigenstates are non-dispersive in
the Brillouin zone these yield the same result.

\begin{figure}
  \centering
  \includegraphics{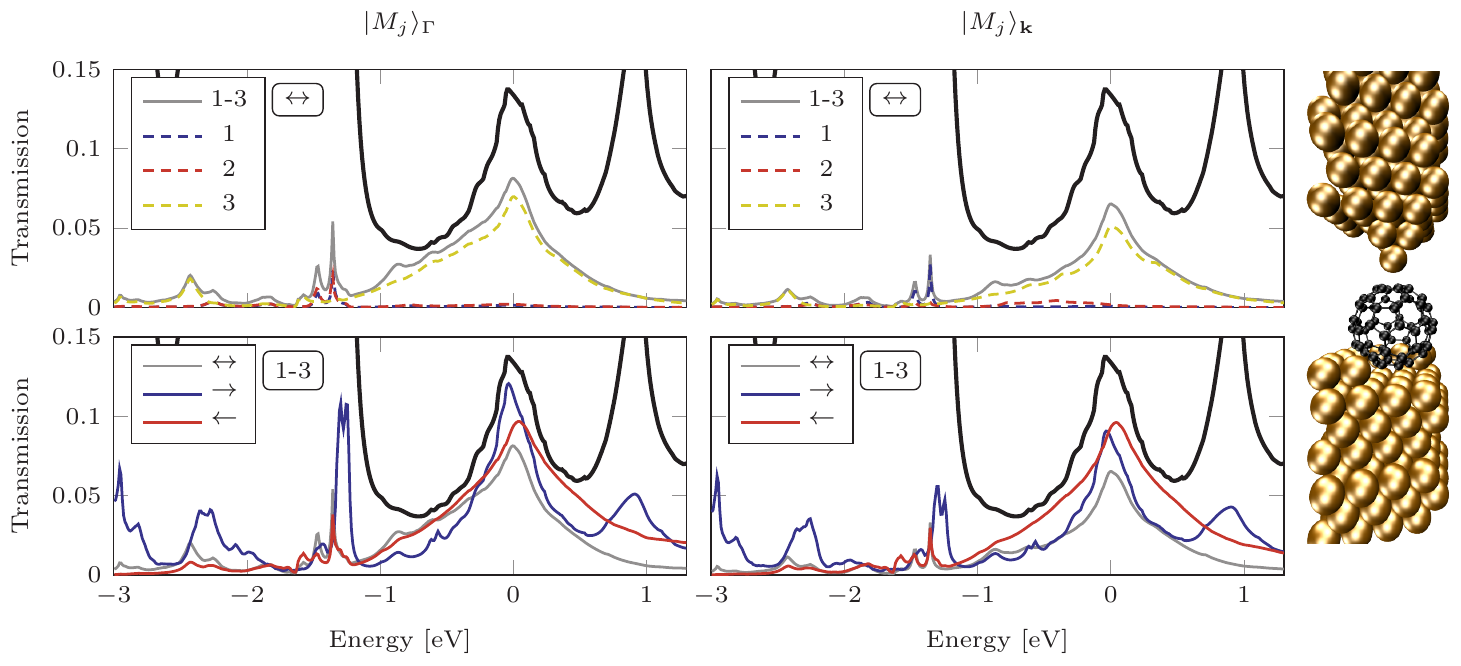}
  \caption{Projected transmission spectra onto the molecular orbitals at zero bias for a Cu(111) surface covered with \Csix\ molecules ($E_F=0$~eV). The full black line is the $\kk$-sampled total transmission (same curve in all panels).
      Projections of the essentially 3-fold degenerate LUMO levels
      are shown with both non-dispersive (left) and dispersive (right) projectors. The
      in-out projectors are used in the top row, while a comparison of the in/out/in-out
      projectors are shown in the bottom row. The Brillouin zone dispersion requires the
      projections to be $\kk$-resolved and it is seen that only a single LUMO MPSH is
      responsible for the majority of the transmission. 
      \label{fig:tbt:proj-c60}
  }
\end{figure}

As an example of projections based on a realistic DFT-NEGF calculation featuring the
$\kk$-dependence we consider again transport through a monolayer of \Csix\ molecules as
shown in \fref{fig:tbt:proj-c60} (setup from Ref.~\cite{Schneider2015}). Due to the
densely packed monolayer coverage of \Csix\ -- one molecule per
    $4\times 4$-surface area of Cu(111) -- there exists a slight dispersion in the
Brillouin zone ($\sim\unitr{10}{meV}$). The transmissions are calculated on a
$13\smash\times13$ Monkhorst-Pack grid \cite{Monkhorst1976}. Here we limit the projections
to involve the three (almost) degenerate MPSH orbitals close to the Fermi energy $E_F$,
i.e., essentially to the lowest unoccupied molecular orbitals (LUMO) of \Csix. The highest
occupied molecular orbitals (HOMO) are located about $\unitr{-1}{eV}$ below $E_F$ while
the LUMO$+1$ are about $\unitr{1}{eV}$ above $E_F$. In \fref{fig:tbt:proj-c60} we compare
the full transmission (thick black) with the projected transmissions. Figures in the left
column are the $\Gamma$-point projectors used in the entire Brillouin zone, while the
right column figures are using the projectors at the given $\kk$ in the Brillouin
zone. Projections labeled by a single integer are individual MPSH projections, while
$1-3$ is the summed projection over all LUMO levels.

The top row of \fref{fig:tbt:proj-c60} shows the projected transmissions using the
projectors as in \eref{eq:tb:proj-IO}. The relative contributions from the dashed lines
clearly reveal that a single LUMO orbital is responsible for carrying the majority of the
transmission, while the other two LUMO orbitals have negligible contribution. The full
projectors ($1-3$) only slightly increases the total transmission compared to the 3rd
projection. The ordering of the levels are according to the energy levels. The total
transmission is not reached due to hybridization of the molecule with the electrode. One
can also observe the effect of intra-molecular coupling yielding a dispersion in the
densely packed monolayer by comparing the results with the $\Gamma$-point versus the
$\kk$-resolved projectors. The latter projection increases the Lorentzian shape of the
transmission peak as well as decrease the hybridized contributions in the energy range
corresponding to the HOMO levels, as expected. 

% 
% PLOT LUMO ORBITAL!!!
% \revision{TF: What is the characteristics
%    of the conducting LUMO orbital as compared with the two non-conducting ones? Could be
%    nice to plot the orbital... Is there a simple, intuitive reason why one couples well?}

The projectors may take either of the forms shown in Eqs.~\eqref{eq:tb:proj}. A comparison
between these different choices are shown in the lower row of \fref{fig:tbt:proj-c60}. The
hybridization of the \Csix\ molecules with the Cu states is strong as reflected in the
many small peaks with the $\rightarrow$ projectors. Contrary, $\leftarrow$ which
projects the molecular orbitals through the weakly coupled tip retains the LUMO energy
range as well as reducing the Brillouin zone dispersion. The latter proves that we have a
small coupling between tips in different supercells resulting in a low dispersion.  It can
also be seen that the total transmission is better retained when using only $\rightarrow$
or $\leftarrow$ which infers a mixing of the scattering states with the molecular
orbitals.

%Note that if the summed projector transmission coincides with a single projector
%transmission it does not necessarily follow that $\bra i\Scat\ket j$ is diagonal. However,
%$\bra i\Scat\ket j\neq\delta_{ij}$ does follow if their individual projected transmissions
%does not sum to the summed projector transmission.
%

\subsection{Phonon transport --- \phtrans}
\label{ssec:phtrans}

The tailoring of heat transport in nanostructures (e.g., via nanostructuring of materials) is
a topic with a large and growing community following the trend of electronic
transport \cite{Wang2008,Dhar2008}. It has promising applications in thermal management and
thermoelectrics, and has even been envisioned for information processing with devices, akin
of electronics, known as ``phononics'' \cite{Li2012}. The transport of heat via phonons can be treated using the Landauer formula 
by replacing (from the electronic case) Fermi-distributions with Bose-Einstein
distributions \cite{Zhang2007,Yamamoto2006}, carrier charge ($e$) with phonon energy
($\hbar\omega$), and using the phonon transmission function $\Xi(\omega)$. The
phonon thermal conductance between two reservoirs at tempertures $T$ and $T+\mathrm dT$,
can thus be calculated as
\begin{equation}
  \kappa_{\mathrm{ph}}(T) = \frac{\hbar^2}{2\pi k_B T^2}\int_{0}^{\infty} \mathrm d\omega\,\omega^2\,\Xi(\omega)\,
  \frac{e^{\hbar\omega/k_BT}}{(e^{\hbar\omega/k_BT}-1)^2}	
  . \label{eq:ThermalConductance}
\end{equation}
The expression for the electronic transmission function in \sref{sec:tbtrans} is easily converted to that of phonons by
replacing the dynamical matrix for the Hamiltonian, using unity as overlap, and the energy
replacement, $\varepsilon+i\eta \rightarrow \omega^2 +i\eta^2$. Recall that all \tbtrans\
functionality may be used in \phtrans, including $N_\idxE\ge1$ terminals. Phonon transport
using the Green function formalism can thus be written as 
\begin{align}
  \G_\qq(\omega)&=\big[(\omega^2+i\eta^2)\mathbf I -
  \Dyn_\qq-\SE_\qq(\omega)\big]^{-1},
  \\
  \Xi_{\idxE\idxE',\qq}(\omega)&=\Tr\big[\G_\qq(\omega)\Scat_{\idxE,\qq}\G^\dagger_\qq(\omega)\Scat_{\idxE',\qq}\big],
\end{align}
with $\Dyn_\qq$ being the dynamical matrix at the $\qq$-point in reciprocal space. %
Currently \sisl\ allows for the extraction of dynamical matrices directly from
\gulp\ \cite{Gale2003} and outputs files to \phtrans\ compatible data format. This enables the
calculation of phonon transport properties for very large systems from empirical
potentials using 3rd party tools.

\begin{figure}
  \centering
  \includegraphics{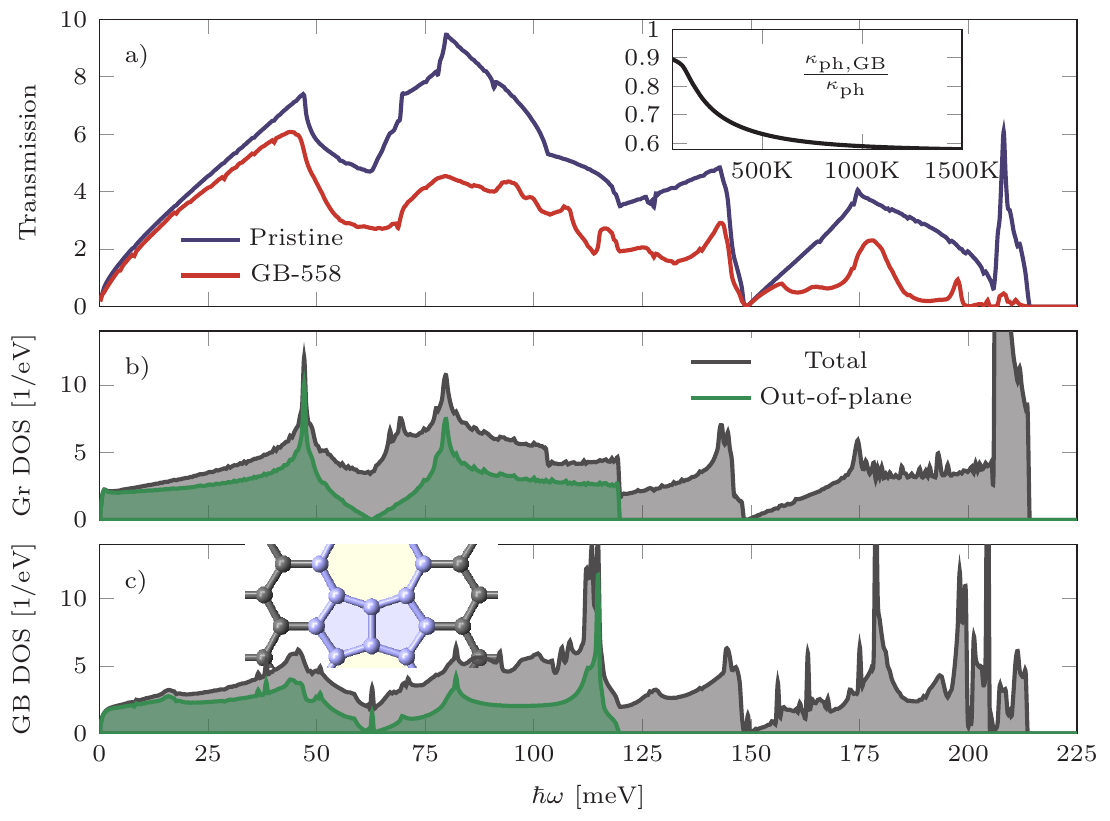}
  \caption{a) Graphene and GB558 (Stone-Wales defect) transmission, $\qq$-averaged, for the
      primitive unit-cell. 
      Insert: Ratio of thermal transmissions as a function of temperature calculated using
      \eref{eq:ThermalConductance}. 
      b) Total DOS and out-of-plane projected DOS per atom of pristine graphene.
      c) Total DOS and projected DOS on out-of-plane phonons, projected onto GB-558 atoms.
      Insert: GB558 structure and projection atoms (blue).
      \label{fig:pht:transport}
  }
\end{figure}

As an example of the capabilities of \phtrans\ we investigated phonon transport in
graphene as well as through a grain boundary. Grain boundaries play a significant role for
both electronic and heat transport in graphene and is an area of intense research
\cite{Huang2011j,Yasaei2015}. Figure \ref{fig:pht:transport} shows the phonon properties
of pristine graphene and the zero-angle grain boundary (GB558, see inset to
\fref{fig:pht:transport}c) for which the electronic transport has previously been studied
\cite{Rodrigues2013}. The quantities are calculated using the Brenner potential
\cite{Brenner2002} in \gulp. The transmission of pristine graphene versus GB558 is
compared in \fref{fig:pht:transport}a. It is seen how the grain-boundary effectively
scatters the phonons and reduces the heat transport to $\sim60$\% at 600K compared to the
pristine case (inset \fref{fig:pht:transport}a).

Figures~\ref{fig:pht:transport}b,c compare atom-resolved phonon DOS inside pristine
graphene and in the grain-boundary, respectively. The grain-boundary hosts quite localized
out-of-plane modes around $\hbar\omega=\unitr{115}{meV}$ and in-plane modes around
$\hbar\omega=\unitr{200}{meV}$ as seen by the peaks in the projected DOS.

\section{Conclusions}
\label{sec:conclusion}

We have presented the Green function technique and its implementation at the DFT-NEGF
level with a generalization of the equations to cover both equilibrium single-electrode
($N_\idxE = 1$) and non-equilibrium multi-electrode ($N_\idxE > 1$) calculations.  The
first case enables equilibrium surface calculations without resorting to slab-approximations,
while the latter case makes studies of non-equilibrium thermo-electric effects
possible using independent electrode Fermi distributions (both chemical potentials \emph{and} temperatures).  The methods are made available in the GPL licensed DFT-NEGF code \tsiesta\ together with the post-processing electron (phonon) transport code
\tbtrans\ (\phtrans). Both codes were re-implemented for extended functionality and optimization.

We now summarize the specific major improvements to the method.  We have implemented
several schemes for equilibrium contour integration to obtain the density matrix. These
improve the convergence properties while reducing the number of integration abscissa in
the equilibrium contour.  Along these lines we generalized the original weighting scheme
for the density matrix integrals to the multi-electrode case ($N_\idxE>2$). Besides the
multi-electrode capabilities we have also implemented a flexible way to include
charge/Hartree electrostatic gate geometries in the DFT-NEGF device region. This enables
investigating gate effects which are ubiquitous for e.g. simulations of functional electronic devices.

The algorithm for calculating the Green function using matrix inversion is crucial for the
performance of any DFT-NEGF code. We have compared 3 implemented methods, LAPACK, MUMPS,
and BTD inversion. We found that the BDT method performs the best and devised a highly
efficient, memory-wise and performance-wise, NEGF variant.  The BTD method relies
critically on the bandwidth of the matrix to be inverted and to accommodate this, we
implemented a variety of pivoting algorithms to reduce bandwidth, memory consumption and
increase performance of the NEGF calculation.  We have furthermore implemented an
efficient method to calculate the spectral function using an efficient propagation
algorithm. Altogether, the performance of \tsiesta\ has improved drastically, compared to
version 3.2, and for one test system an impressive $\sim100$ times speed-up was achieved.
OpenMP 3.1 threading was also implemented in \tsiesta\ which allows to reach unprecedented
system sizes with the DFT-NEGF method.

As a post-processing tool \tbtrans\ has been presented for calculation of the transmission
function from any Hamiltonian or dynamical matrix (\phtrans, phonon-transport). Our
implementation involves a new separation algorithm where the transport properties --- as
well as local quantities such as DOS, bond-currents, etc. --- may be efficiently calculated
in a selected subspace of the device cell. A recurring question within molecular
electronics is the origin of the electron carrier orbital. We presented a novel method to
calculate projected transmissions using either single molecular orbitals or a combination
of several orbitals. The projection method includes $\Gamma$ and $\kk$ point projectors.

Besides the many optimizations mentioned above we also implemented an efficient interpolation
method of finite-bias Hamiltonians for \tbtrans. This reduces the computational burden of
full $I-V$ curves. We presented linear and spline interpolation and demonstrated how spline
interpolation proves very accurate, even for complex systems.
\tbtrans\ is now featured as a stand-alone transport code which enables other codes to
interface to it. In particular, we developed \sisl\ which is a generic Python code for
creating and manipulating Hamiltonians. \sisl\ can already now interact with several codes
as well as extracting the dynamical matrix from \gulp\ and passing it to \phtrans.

All together \tsiesta\ and \tbtrans\ (and its offshoot \phtrans) now utilize highly
scalable and efficient algorithms. Both codes fully implement $N_\idxE\ge1$ electrode
Green function techniques with full customization of each electrode. Our novel
implementations have thus enabled everyday DFT-NEGF (tight-binding) calculations in excess
of 10,000 (1,000,000) orbitals which earlier would have seemed insurmountable.

%\revision{MB: maybe we need to do a large killer calculation to prove this point.. could
%    run while it is with referees...}

\section{Acknowledgments}

\iffalse % Long

We thank Mads Engelund, Pedro Brandimarte and Georg Huhs for testing and feedback on the
software. The DIPC funded an external stay for NP during code development. The Danish
e-Infrastructure Cooperation (DeIC) for provided computer resources. The Center for
Nanostructured Graphene (CNG) is sponsored by the Danish Research Foundation, Project
DNRF103.
NP acknowledges financial support from EU H2020 project no.~676598, ``MaX: Materials at
the eXascale'' Center of Excellence in Supercomputing Applications.
TF acknowledges financial support from the Basque Departamento de Educaci\'on and the
UPV/EHU (Grant No.~IT-756-13), the Spanish Ministerio de Econom\'{\i}a y Competitividad
(Grant No.~MAT2013-46593-C6-2-P), and the European Union FP7-ICT project PAMS (Contract
No.~610446).
NL acknowledges financial support from the Spanish Ministerio de Econom\'{\i}a y
Competitividad (Grant No.~MAT2015-66888-C3-2-R).
AG acknowledges financial support from the Spanish Ministry of Economy and
Competitiveness, through the ``Severo Ochoa'' Program for Centers of Excellence in R\&D
(SEV- 2015-0496) and through grants FIS2012-37549-C05-05 and FIS2015-64886-C5-4-P, and
from Generalitat de Catalunya (2014 SGR 301).

\else % Short

We thank Mads Engelund, Pedro Brandimarte and Georg Huhs for testing and feedback on the
software. The DIPC funded an external stay for NP during code development. The
Danish e-Infrastructure Cooperation (DeIC) provided computer resources. The Center for
Nanostructured Graphene (CNG) is sponsored by the Danish Research Foundation, Project
DNRF103.  We acknowledge financial support from EU H2020 project no.~676598, ``MaX:
Materials at the eXascale'' Center of Excellence in Supercomputing Applications, the
Basque Departamento de Educaci\'on and the UPV/EHU (IT-756-13), the Spanish Ministerio de
Econom\'{\i}a y Competitividad (MAT2013-46593-C6-2-P, MAT2015-66888-C3-2-R,
FIS2012-37549-C05-05, and FIS2015-64886-C5-4-P as well as the ``Severo Ochoa'' Program for
Centers of Excellence in R\&D SEV-2015-0496), the European Union FP7-ICT project PAMS
(Contract No.~610446), and the Generalitat de Catalunya (2014 SGR 301).

\fi

\end{document}